\documentclass{IEEEtran}

\usepackage{cite} \usepackage[cmex10]{amsmath} \usepackage{graphicx}
\usepackage[caption=false,font=footnotesize]{subfig}
\usepackage{amssymb} \usepackage{accents} \usepackage{multirow}
\usepackage{algorithmic}
\usepackage{algorithm}

\DeclareMathOperator{\minimize}{minimize}
\DeclareMathOperator{\maximize}{maximize}
\newtheorem{prop}{Proposition} 
\newtheorem{myremark}{Remark}

\newsavebox{\ieeealgbox}
\newenvironment{boxedalgorithmic}
  {\begin{lrbox}{\ieeealgbox}
   \begin{minipage}{\dimexpr\columnwidth-2\fboxsep-2\fboxrule}
   \begin{algorithmic}[1]}
  {\end{algorithmic}
   \end{minipage}
   \end{lrbox}\noindent\fbox{\usebox{\ieeealgbox}}}

\allowdisplaybreaks

\begin{document}

\title{Energy Management in Heterogeneous Networks with Cell
  Activation, User Association and Interference Coordination }

\author{Quan Kuang, and  Wolfgang Utschick,~\IEEEmembership{Senior Member,~IEEE}%
\thanks{This is an extended version of a paper to appear in IEEE Transactions on Wireless Communications (TWC) with the same title, containing all detailed proofs. Manuscript submitted to IEEE TWC July 15, 2015; revised November 25, 2015; accepted January 29, 2016 Part of the work has been accepted to present at ICASSP 2016 \cite{kuangICASSP16}.}%
\thanks{Q. Kuang was with the Department of Electrical
and Computer Engineering, Technische Universit\"{a}t M\"{u}nchen,
    Munich 80333, Germany. He is now with Panasonic R\&D Center Germany (e-mail: quan.kuang@eu.panasonic.com).
 W. Utschick is with the Department of Electrical
and Computer Engineering, Technische Universit\"{a}t M\"{u}nchen,
    Munich 80333, Germany.}}

\maketitle

\begin{abstract}

  The densification and expansion of wireless network pose new
  challenges on interference management and reducing energy
  consumption. This paper studies energy-efficient resource management
  in heterogeneous networks by jointly optimizing cell activation,
  user association and multicell multiuser channel assignment,
  according to the long-term average traffic and channel
  conditions. The proposed framework is built on characterizing the
  interference coupling by pre-defined interference patterns, and
  performing resource allocation among these patterns. In this way,
  the interference fluctuation caused by (de)activating cells is
  explicitly taken into account when calculating the user achievable
  rates. A tailored algorithm is developed to solve the formulated problem in
  the dual domain by exploiting the problem structure, which gives a
  significant complexity saving. Numerical results show a huge improvement in energy saving achieved
  by the proposed scheme. The user association derived from the
  proposed joint resource optimization is mapped to standard-compliant
  cell selection biasing. This mapping reveals that
  the cell-specific biasing for energy saving is quite different from
  that for load balancing investigated in the literature.

\end{abstract}


\section{Introduction}

There exists an emerging paradigm shift in wireless infrastructure
systems, where densely deployed small and low-cost base stations (BSs)
are embedded into the conventional cellular network topology to form a
so-called heterogeneous network (HetNet) \cite{Bhushan2014}. In a
dense HetNet, BSs are typically deployed to satisfy the peak traffic
volume and they are expected to have low activity outside rush hours
such as nighttime. Hence, there is a high potential for energy saving
if BSs can be switched off according to the traffic load
\cite{Son2011a}.

Obviously, cell activation is coupled with user association: the
users in the muted cells must be re-associated with other BSs. In
addition, cell muting and user re-association impose further
challenges on interference management, since the user may not be
connected to the BS with the strongest signal strength. This
interference issue can be resolved by resource coordination, i.e.,
properly sharing the channels among multiple cells and then
distributing them to the associated users in each cell. Hence, to
achieve energy efficiency, multicell multiuser channel assignment
should be integrated into the optimization of the cell activation and
user association.

However, the resource management that considers the above elements
jointly is very challenging mathematically because the inter-cell
interference coupling leads to the inherent non-convexity in the
optimization problems. To make the problems tractable, the previous
studies relied on worst-case interference assumption
\cite{Pollakis2012,Cavalcante2014}, average interference assumption
\cite{Son2011a,Kim2013a}, or neglecting inter-cell interference
\cite{Su2013}. In these works, the interference was assumed
\emph{static} (or absent), i.e., independent of the resource
allocation decisions in each cell when estimating the user achievable
rate. Clearly, this is a suboptimal design because the BS deactivation
will cause interference fluctuation in the network, hence affecting
the user rate.

This paper develops a new framework for energy-efficient resource
management to consider the coupling effect of the inter-cell
interference caused by cell activation. The idea is to pre-calculate
the user rate under each possible \emph{interference pattern} (i.e. an
interference scenario in the network, described as one combination of
ON/OFF activities of the BSs), and then perform resource allocation
among these patterns. This allocation yields the actual interference
and the corresponding user achievable rates that well match the
interference at the same time.

Other related works include \cite{Liao2013,YuanMingTWC,ShixinTWC2015}
and references therein, where the sparse optimization techniques
similar to one adopted in this paper have been used to optimize BS
activation and/or coordinated multi-point processing (CoMP) according
to the instantaneous channel state condition, with pre-determined
channel allocation. By contrast, we consider a slow adaptive strategy
over a period of many minutes to reduce the control
overhead. Moreover, channel allocation is introduced as an effective
way for interference mitigation.

The system model and problem formulation in this work stem from
our prior studies in \cite{Kuang2014a,kuangISWCS}, where multi-pattern
resource allocation were exploited to maximize the network proportional
fairness utility with a fixed number of active cells. The similar idea
of allocating spectrum among multiple reuse patterns in a slow timescale
has also been adopted in independent works \cite{Zhuangtobepublished, ZhuangEnergy} for
optimizing queueing delay and/or energy management. Nevertheless, in this paper a unique dual cutting plane approach is used to develop efficient solving algorithms
(see a summary of contributions below).

%

The main contributions of this paper are:
\begin{itemize}
\item[1)] An energy-efficient resource management problem is proposed
  to serve the user demand with minimum network power
  consumption. It jointly optimizes the cell activation, user
  association, and multicell multiuser channel assignment, according
  to the long-term average traffic and channel conditions. The
  interference coupling caused by cell activation is taken into
  account when estimating the user achievable rates, which is done by
  pre-defining interference patterns. Although a network of $B$ cells
  has $2^B$ possible interference patterns, we show that a
  small number of patterns are sufficient to obtain accurate estimates
  of the user rates.

\item[2)] A tailored algorithm is developed for solving the formulated
  resource management problem. There are two ingredients in this
  development: the reweighted $\ell_1$ minimization \cite{Candes2008}
  is used to tackle the $\ell_0$ term in the objective, and a cutting
  plane method is used for solving the dual problem by exploiting the
  problem structure, resulting in a significant complexity saving
  compared to directly applying standard interior-point solvers. This
  complexity reduction makes it possible to include all $2^B$ patterns
  in the optimization problem for reasonably-sized networks, providing
  an energy saving benchmark for comparison with other schemes where
  pattern selection is restricted.

\item[3)] Using the proposed framework, existing resource management
  proposals are evaluated and compared in a unified manner, where the
  interference coordination is either involved or not, and the user
  association is either optimized jointly with the resource allocation
  or performed by simple cell selection biasing. In this way, the
  impacts of interference coordination and user association on energy
  saving are individually characterized.

\item[4)] The user association decision obtained by the proposed joint
  resource optimization is mapped to standard-compliant cell selection
  biasing \cite{Damnjanovic2011}. This mapping reveals that, in
  contrast to previous studies (e.g., \cite{Ye2013}) where common
  per-tier biasing is sufficient in terms of load balancing and
  network rate utility maximization, the energy-efficient solution
  requires individual biasing for different cells in the same tier.

\end{itemize}

The remainder of this paper is organized as follows. Section
\ref{sec:system-model} introduces the system model. Rate-constrained
energy saving problem is formulated and studied in Section
\ref{sec:rate-constr-energy}. Section \ref{sec:unif-study-baseline}
develops a unified view on a wide range of existing resource
management schemes, and the performance comparison is provided in
Section \ref{sec:results} followed by conclusion in Section \ref{sec:conclusion}.

\section{System model}
\label{sec:system-model}


We consider a downlink HetNet, where a number of small cells\footnote{Cell and BS are used
  interchangeable in this paper} are
embedded in the conventional macro cellular network.  The set of all
(macro and small) cells is denoted by $\mathcal{B} = \{1,2,\cdots,B
\}$. The cells can be switched on or off every time period $T$ (say,
many minutes) according to the fluctuations in the traffic load. We
select $K$ test points in the network as the representation of typical
user locations \cite{Cavalcante2014}, denoted by $\mathcal{K} =
\{1,2,,\cdots K\}$.  The user demand of each test point $k \in
\mathcal{K}$ is represented by a minimum required average rate $d_k$,
which is assumed known from traffic estimation and user QoS
requirements (see Section \ref{sec:modelingIssue} for more discussion
on test point selection and demand modeling). We are interested in
developing adaptive strategies for every period of $T$ to accommodate
the user demand with minimum network energy consumption, taking into
account the inter-cell interference coupling.

The enabling mechanism is to characterize the interference by
specifying the interference patterns, each of which defines a
particular ON/OFF combination of BSs. We use the pattern activity
vector $\mathbf{a}_i = (a_{i1}, a_{i2}, \cdots, a_{iB})^T$ to
indicate the ON/OFF activity of the BSs under pattern $i$, where
\begin{equation}\label{pattern_actVect}
  a_{ib} =    \left\{  \begin{array}{rl}
      1 & \text{if BS $b$ is ON under pattern $i$}   \\
      0 & \text{otherwise}
    \end{array} \right.
\end{equation}
We denote the set of pre-defined patterns by $\mathcal{I} =
\{1,2,\cdots, I\}$ and further define the matrix $\mathbf{A} =
(\mathbf{a}_1, \mathbf{a}_2,\cdots, \mathbf{a}_I)$ to
combine the activity vectors for all candidate patterns. In order to
fully characterize the interference scenarios in a network of $B$
cells, generally speaking, $2^B$ patterns are needed. However, since
BSs located far away have weak mutual interference, omitting some
patterns will not affect the accurate estimation of user achievable
rates. We will discuss more on this next (see Proposition 1 and
Section \ref{sec:cell-clust-activ}).

\begin{figure}[!t]
  \centering
   \includegraphics[width=3.2In]{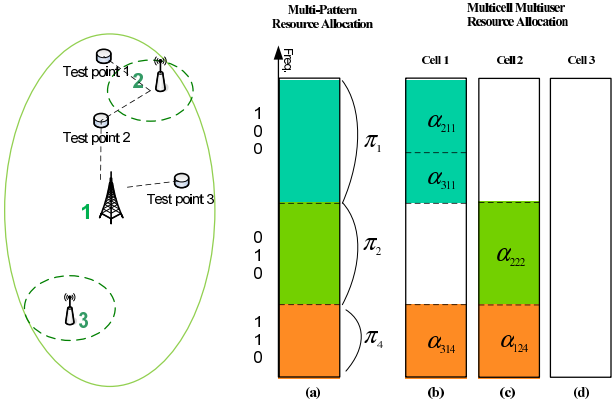}
  \caption{An illustration of multi-pattern resource allocation in a
    HetNet consisting of one macro (Cell 1), two small cells (Cells 2
    and 3), and 3 test points.
    Bar chart (a) gives a result of spectrum allocation among all 7
    patterns, where $\pi_i$ is the fraction allocated to pattern
    $i$. As shown, $\pi_3 = \pi_5 = \pi_6 = \pi_7 = 0$. Result of (a)
    can be directly translated into spectrum allocation among
    different cells as shown in (b), (c), and (d) respectively, where
    forbidden frequency resources are indicated in white
    color. $\alpha_{kbi}$ denotes the fraction allocated to test point
    $k$ by BS $b$ under pattern $i$. In the shown result, Cell 1
    serves test points 2 and 3, and Cell 2 serves points 1 and 2.
  }\label{fig1_illustration}
\end{figure}

Fig.~\ref{fig1_illustration} illustrates the idea of multi-pattern
formulation. Firstly, the multi-cell channel allocation is translated
into partitioning the spectrum across all patterns.
In a slow timescale considered in this paper, all frequency resources
can be assumed to have equal channel conditions. Denote the spectrum
allocation profile by $\boldsymbol{\pi} = (\pi_1, \ldots, \pi_i,
\ldots, \pi_I)^T \in \Pi$,
where $\pi_i$ represents the fraction of the total bandwidth allocated to
pattern $i$ and $\Pi = \{\boldsymbol{\pi} : \sum_{i\in\mathcal{I}} \pi_i = 1, \pi_i
\geq 0, \forall i\}$. Then the total bandwidth fraction allocated to BS $b$ is
$\mathbf{A}_{b\bullet}\times \boldsymbol{\pi}$, where $\mathbf{A}_{b\bullet}$
is the $b$-th row of the matrix $\mathbf{A}$.

Secondly, test point association and multiuser channel allocation can
also be easily done thanks to the multi-pattern formulation. In more
detail, denote by $\alpha_{kbi} \geq 0$ the fraction of resources that
BS $b$ allocates to test point $k$ under pattern $i$. Naturally, each
BS is allowed to use up to $\pi_i$ resources under pattern $i$ for its
associated test points, expressed as $\sum_{k\in \mathcal{K}}
\alpha_{kbi} \leq \pi_i, \forall b, \forall i$. Note that the
association is implicitly indicated by $\alpha_{kbi}$, i.e.,
$\alpha_{kbi} > 0$ means test point $k$ is associated with BS $b$
under pattern $i$, while zero value of $\alpha_{kbi}$ means that they
are not connected.

Finally, we define the usage of BS $b$ as $\rho_b=\sum_k \sum_i
\alpha_{kbi}$, with $0 \leq \rho_b \leq 1, \forall b \in
\mathcal{B}$. Then cell $b$ is active if $\rho_b$ is nonzero.


\subsection{Rate model}

Assuming flat power spectral density (PSD) of BS transmit power and
the noise, the received SINR of the link connecting BS
$b$ to test point $k$ under pattern $i$ is
\begin{equation}\label{SINR}
  \textrm{SINR}_{kbi} = \frac{a_{ib} P_b G_{bk}\|h_{bk,n}\|^2 }{\sigma^2 + \sum_{l\neq b} a_{il} P_l G_{lk}\|h_{lk,n}\|^2 }
\end{equation}
where $a_{ib}$ is the cell activation indicator as given in
(\ref{pattern_actVect}), $P_b$ is the PSD of BS $b$, $\sigma^2$ is the
received noise PSD. We denote the channel gain between BS $b$ and test
point $k$ over the $n$-th frequency resource by $\sqrt{G_{bk}}
h_{bk,n}$ where $G_{bk}$ is the large-scale coefficient including
antenna gain, path loss and shadowing, and $h_{bk,n}$ accounts for the
small-scale fading. We assume $\{h_{bk,n},\forall b, \forall k,
\forall n\}$ are independent and identically distributed
(i.i.d.). Hence, the ergodic rate of test point $k$ served by the
$b$-th BS under pattern $i$ can be written as
\begin{equation}\label{RatePerCarrier}
  \bar{r}_{kbi}  = \alpha_{kbi} \underbrace{ W
    \mathbb{E}_{\mathbf{h}} \left[\log_2 \left(1+
        \textrm{SINR}_{kbi}\right) \right]}_{ \triangleq r_{kbi}}  = \alpha_{kbi} r_{kbi}  
\end{equation}
where $W$ is the system bandwidth, $\alpha_{kbi}$ denotes the fraction
of bandwidth that BS $b$ allocates to test point $k$ under pattern $i$,
$\mathbf{h} \triangleq (h_{1k,n},h_{2k,n},\ldots, h_{Bk,n})$.

Finally, the total rate of test point $k$ is obtained as
\begin{equation}\label{RateCombined}
  R_k  = \sum_{i \in \mathcal{I}} \sum_{b \in \mathcal{B}} \bar{r}_{kbi}= \sum_{i \in \mathcal{I}} \sum_{b \in \mathcal{B}} \alpha_{kbi} r_{kbi}.
\end{equation}
Note that $r_{kbi}$ can be pre-calculated using (\ref{RatePerCarrier})
and hence treated as constants during the resource optimization. In
\eqref{RateCombined}, single-BS association restriction is not
enforced, i.e., test point $k$ is allowed to be connected to multiple
BSs. The physical interpretation of this relaxation could be that the
signals for test point $k$ from multiple BSs are encoded and decoded
separately by treating the signals from all BSs except one as
interference (see \eqref{SINR}), and then the rate contributions from
multiple BSs are summed up to give the final rate of test point
$k$. We will show in Section \ref{sec:performanceProposed} that this
relaxation turns out to be really tight, in the sense that almost all
of the test points are associated with single BS as the result of
optimization.

\subsection{Energy consumption model}
As mentioned previously, the BS usage is defined as
$\rho_b =
\sum_k\sum_i \alpha_{kbi}$, $\forall b$. A typical power consumption model for BSs
consists of two types of power consumption: fixed power consumption
and dynamic power consumption that is proportional to BS's utilization
\cite{Son2011a}. Denote by $P_b^{\text{OP}}$ the maximum
operational power of BS $b$ if it is fully utilized (i.e., $\rho_b =
1$), which includes power consumption for transmit antennas as well as
power amplifier, cooling equipment and so on. We can then express the
total power consumption by all BSs as
\begin{IEEEeqnarray}{rCl}
  \label{eq:19}
P^{\text{tot}} =  \sum_{b\in \mathcal{B}} \left[ (1-q_b)\rho_b P_b^\textrm{OP} +
    q_b |\rho_b|_0 P_b^\textrm{OP} \right]
\end{IEEEeqnarray}
where $q_b \in (0,1]$ is the portion of the fixed power
consumption for BS $b$ as long as it is switched on, and $|x|_0$ is the
function that takes the value of 0 if $x=0$ or the value 1
otherwise (i.e., $\ell_0$-norm applied to a scalar). Note that by setting $q_b = 1$ we arrive at a constant
energy consumption model considered in
\cite{Pollakis2012,Niu2010}, which is a reasonable assumption for
macro BSs. However, the small BSs such as pico or femto BSs may have smaller
value of $q_b$ because they do not usually have a big power amplifier
or cooling equipment.

\section{Rate-constrained energy saving}
\label{sec:rate-constr-energy}

\subsection{Problem formulation}
The joint optimization of cell activation, user association and
interference coordination via channel assignment for network energy
saving can be formulated as
\begin{IEEEeqnarray}{rCl}\label{P1_rateConstrained}
\IEEEyesnumber\IEEEyessubnumber*
  \displaystyle\mathop{\minimize}_{\boldsymbol{\alpha},\boldsymbol{\pi}}
  \quad
  &&    P^{\text{tot}}= \sum_{b\in \mathcal{B}} \left[ (1-q_b)\rho_b P_b^\textrm{OP} +
    q_b |\rho_b|_0 P_b^\textrm{OP} \right]  \label{obj} \IEEEeqnarraynumspace\\
  \text{subject to} \quad && \rho_b =  \sum_{k\in\mathcal{K}}
  \sum_{i\in\mathcal{I}} \alpha_{kbi}, \  \forall b \label{cons_rho} \\
  && \sum_{i \in \mathcal{I}} \sum_{b \in \mathcal{B}}  \alpha_{kbi} r_{kbi} \geq d_k, \ \forall k  \label{conQoS} \\
  && \sum_{k \in \mathcal{K}} \alpha_{kbi} \leq \pi_i, \forall b,\ \forall i    \label{const_BS allo}\\
  && \sum_{i \in \mathcal{I}} \pi_i = 1  \label{cons_pi} \\
  && \pi_i \geq 0, \ \forall i, \quad \alpha_{kbi} \geq 0, \ \forall
  k,b,i  \label{con_nonnegative}
\end{IEEEeqnarray}
where (\ref{conQoS}) specifies the user demand of all test points, and
all variables and parameters have been explained in Section \ref{sec:system-model}.

The difficulty of solving \eqref{P1_rateConstrained} lies in two
facts. The first is the combinatorial objective function involving the
$\ell_0$-norm. The other is that the number of all possible patterns
in the network grows exponentially with the number of cells as $2^B$,
resulting in huge problem dimension for a moderate-sized
network. Fortunately, the following Proposition \ref{prop0} identifies
that only a small number of patterns out of $2^B$ are needed for
resource allocation to achieve the optimality.

\begin{prop}\label{prop0}
  There exists an optimal solution to problem
  \eqref{P1_rateConstrained} that activates at most $K+B+1$ patterns,
  i.e., $|\{i\in \mathcal{I} : \pi_i > 0 \}| \leq K+B+1$, where $K$ and $B$ are the
  number of users and number of cells in the network, respectively.
\end{prop}

\begin{IEEEproof}
The proof is given in Appendix.
\end{IEEEproof}

Proposition \ref{prop0} indicates that if we know the set of most
important patterns beforehand, the complexity of solving the resource
allocation problem \eqref{P1_rateConstrained} can be significantly
reduced by restricting the candidate patterns to this set. Section \ref{sec:cell-clust-activ}
suggests a practical guideline for pre-selecting candidate patterns. The effectiveness
of this selection criterion in terms of energy saving will be evaluated in Section
\ref{sec:results}.

\subsection{Feasibility test}\label{sectionFeasiTest}

Before describing the proposed method to solve
(\ref{P1_rateConstrained}), we introduce a rate balancing scheme to
test the feasibility of (\ref{P1_rateConstrained}) as follows. First,
the rate requirement can be expressed using a normalized vector
$\boldsymbol{\beta} = (\beta_1, \cdots, \beta_K)^T$ where $\beta_k =
\frac{d_k}{\sum_{k\in \mathcal{K}} d_k}$. Then the feasibility of problem
(\ref{P1_rateConstrained}) can be determined by solving the following
rate balancing problem:
\begin{IEEEeqnarray}{rCl}\label{P_rateBalancing}
\IEEEyesnumber\IEEEyessubnumber*
  \displaystyle\mathop{\minimize}_{\boldsymbol{\alpha},\boldsymbol{\pi},
    R_{\textrm{sum}}} \quad
  &&    - R_{\textrm{sum}}   \\
  \text{subject to} \quad
  && \beta_k R_{\textrm{sum}} - \sum_{i \in \mathcal{I}} \sum_{b \in \mathcal{B}}  \alpha_{kbi} r_{kbi} \leq 0 , \ \forall k  \label{con_rateBal_1} \\
  && (\boldsymbol{\alpha},\boldsymbol{\pi}) \in \mathcal{X}
\end{IEEEeqnarray}
where $\mathcal{X}$ is defined by (\ref{const_BS allo}),
(\ref{cons_pi}) and (\ref{con_nonnegative}). Let
$R_{\textrm{sum}}^\star$ denote the optimal value of problem
(\ref{P_rateBalancing}). The original problem
(\ref{P1_rateConstrained}) is feasible if and only if
$R_{\textrm{sum}}^\star \geq \sum_k d_k$.

Note that problem (\ref{P_rateBalancing}) is a linear optimization
problem and always feasible. It can be efficiently solved by, e.g.,
interior-point methods, using off-the-shelf solvers, if the problem
dimension $\mathcal{O}(IKB)$ is small. However, it is also desirable
to solve \eqref{P_rateBalancing} by involving a large number of
patterns. This could happen when we consider all possible $2^B$
patterns in order to calculate an optimal performance benchmark in a
reasonable-sized network, or when the pre-selection still results in
lots of candidate patterns for a large-scale network.  In such case,
the existing interior-point solvers, such as SeDuMi and SDPT3, cannot be applied, since they typically have cubic
computational complexity in the problem dimension \cite{Nesterov1994}.
Fortunately, the problem has an interesting structure that facilitates
a tailored cutting plane method to solve the dual problem.

The dual objective function can be written as
\begin{IEEEeqnarray}{rCl}\label{dualObj_rateBalancing}
  g(\boldsymbol{\lambda}) =
  \inf_{\substack{(\boldsymbol{\alpha},\boldsymbol{\pi})\in \mathcal{X} \\
    R_{\textrm{sum}}}}\left\{ R_{\textrm{sum}}(\sum_k \lambda_k -1) -
    \sum_{k,b,i} \alpha_{kbi} \frac{r_{kbi}\lambda_k}{\beta_k}
  \right\} \IEEEeqnarraynumspace
\end{IEEEeqnarray}
where $\boldsymbol{\lambda} = (\lambda_1, \cdots, \lambda_K)^T$ with
$\lambda_k$ being the multiplier for the $k$-th inequality constraints
in (\ref{con_rateBal_1}). This function is unbounded unless $\sum_k
\lambda_k =1$. Therefore, the corresponding dual problem can be stated
as
\begin{IEEEeqnarray}{rCl}\label{dualprob_rateBl}
  \mathop{\maximize}_{\boldsymbol{\lambda} \in \Lambda} \
  \mathop{\minimize}_{(\boldsymbol{\alpha},\boldsymbol{\pi})\in
    \mathcal{X}} - \sum_{k,b,i} \alpha_{kbi}
  \frac{r_{kbi}\lambda_k}{\beta_k}
\end{IEEEeqnarray}
where $\Lambda = \{\boldsymbol{\lambda} : \sum_k \lambda_k =1,
\lambda_k \geq 0, \forall k \}$. Since the primal problem
(\ref{P_rateBalancing}) is feasible, strong duality holds for this
linear program \cite[Ch.5]{Boyd2004}. So (\ref{P_rateBalancing}) can
be alternatively solved by the dual problem
(\ref{dualprob_rateBl}). The following Proposition \ref{prop1} plays
an important role for developing efficient solving algorithms.
\begin{prop}\label{prop1}
  The inner minimization of problem (\ref{dualprob_rateBl}) with fixed
  $\boldsymbol\lambda$ has a
  closed-form solution, which is
  \begin{equation}\label{solu_alp}
    \bar{\alpha}_{kbi} =    \left\{  \begin{array}{rl}
        1 & \text{if} \ i=\bar{i}, k=\bar{k}(b, \bar{i})
          \\
        0 & \text{otherwise}
      \end{array} \right.
  \end{equation}
  and
  \begin{equation}\label{solu_pi}
    \bar{\pi}_i =    \left\{  \begin{array}{rl}
        1 & \text{if} \ i=\bar{i}  \\
        0 & \text{otherwise}
      \end{array} \right.
  \end{equation}
  where
  \begin{equation}\label{optimal_k}
    \bar{k}(b,i) = \arg \min_k \accentset{\circ}{r}_{kbi}
  \end{equation}

\begin{equation}\label{optimal_i}
  \bar{i} = \arg \min_i \sum_b \accentset{\circ}{r}_{\bar{k}(b,i)bi}
\end{equation}
with
\begin{equation}\label{revised_r}
  \accentset{\circ}{r}_{kbi} =  - \frac{r_{kbi}\lambda_k}{\beta_k}.
\end{equation}
\end{prop}

\begin{IEEEproof}
  The inner minimization of (\ref{dualprob_rateBl}) with respect to
  $(\boldsymbol{\alpha},\boldsymbol{\pi})$ can be rewritten as the
  following inner-outer formulation:
  \begin{IEEEeqnarray}{rCl}\label{inner_outer linear}
    \mathop{\minimize}_{\pi_i \geq 0, \sum_i \pi_i =1} \
    \mathop{\minimize}_{\alpha_{kbi} \geq 0, \sum_k \alpha_{kbi} \leq
      \pi_i} \sum_k \sum_i \sum_b \alpha_{kbi}
    \accentset{\circ}{r}_{kbi}
  \end{IEEEeqnarray}
  where $\accentset{\circ}{r}_{kbi}$ is defined in
  (\ref{revised_r}). Since $\accentset{\circ}{r}_{kbi} \leq 0$, it is
  clear that the inner problem of (\ref{inner_outer linear}) with
  respect to $\boldsymbol{\alpha}$ is solved by each BS exclusively
  allocating maximum allowable resources to the single user who
  benefits the most for each pattern, i.e.,
  \begin{equation}\label{solu_alp_inner}
    \alpha_{kbi} =    \left\{  \begin{array}{rl}
        \pi_i & \text{if} \ k=\bar{k}(b, i)  \\
        0 & \text{otherwise}
      \end{array} \right.
  \end{equation}
  where $\bar{k}$ is expressed as (\ref{optimal_k}). Substituting the
  solution of (\ref{solu_alp_inner}) back to (\ref{inner_outer
    linear}), we arrive at the following problem:
  \begin{IEEEeqnarray}{rCl}\label{outer linear}
    \mathop{\minimize}_{\pi_i \geq 0, \sum_i \pi_i = 1}  \sum_{i}
    \pi_i \sum_{b} \accentset{\circ}{r}_{\bar{k}(b,i)bi}
  \end{IEEEeqnarray}
  which is solved by pooling all resources to one pattern. So we
  obtain the solutions of (\ref{optimal_i}) and (\ref{solu_pi}), hence
  (\ref{solu_alp}).
\end{IEEEproof}

By means of introducing an auxiliary variable $z$, the dual problem
(\ref{dualprob_rateBl}) can be equivalently expressed as
\begin{IEEEeqnarray}{rCl}\label{dualprob_rateBl_re}
\IEEEyesnumber\IEEEyessubnumber*
  \mathop{\maximize}_{\boldsymbol{\lambda} \in \Lambda, z}  \quad &&  z  \\
  \textrm{subject to} \quad && -\sum_{k,b,i} \alpha_{kbi}
  \frac{r_{kbi}\lambda_k}{\beta_k} \geq z, \forall
  (\boldsymbol{\alpha}, \boldsymbol{\pi}) \in \mathcal{X}\label{infCons}
\end{IEEEeqnarray}
which unfortunately has infinitely many constraints in
(\ref{infCons}). The cutting plane algorithm solves an approximation
at every iteration by considering only finite number of constraints
and then refines this approximation by adding more constraints (cuts)
for next iterations. Specifically, the following \emph{master
  problem} is solved during the $l$-th iteration, given
$\boldsymbol{\alpha}^{(0)},\cdots, \boldsymbol{\alpha}^{(l-1)} \in
\mathcal{X}$:
\begin{IEEEeqnarray}{rCl}\label{master_rateBl_dual}
  \mathop{\maximize}_{\boldsymbol{\lambda} \in \Lambda, z}  \quad &&  z \IEEEnonumber \\
  \textrm{subject to} \quad && -  \sum_{k,b,i} \lambda_k
   \frac{\alpha^{(j)}_{kbi} r_{kbi}}{\beta_k} \geq z, \ \forall  j\in \{0,1,\cdots, l-1\}. \IEEEeqnarraynumspace
\end{IEEEeqnarray}
Let $(\boldsymbol{\lambda}^{(l)}, z^{(l)})$ be an optimal solution to
the above problem \eqref{master_rateBl_dual}. Then we have $z^{(l)}
\geq z^\star$, where $z^\star$ is the optimal value of the original
dual problem (\ref{dualprob_rateBl_re}), because the problem
(\ref{master_rateBl_dual}) has less restrictive constraints. In order
to check whether $(\boldsymbol{\lambda}^{(l)}, z^{(l)})$ is also an
optimal solution to the original dual problem, we need to solve the
inner minimization of problem (\ref{dualprob_rateBl}) for given
$\boldsymbol{\lambda}^{(l)}$ :
\begin{IEEEeqnarray}{rCl}\label{subproblem_rateBl}
  \mathop{\minimize}_{(\boldsymbol{\alpha}, \boldsymbol{\pi}) \in
    \mathcal{X}} \quad - \sum_{k,b,i} \alpha_{kbi}
  \frac{r_{kbi}\lambda_k^{(l)}}{\beta_k},
\end{IEEEeqnarray}
i.e., the dual function $g$ is evaluated at
$\boldsymbol{\lambda}^{(l)}$.  Let $(\boldsymbol{\alpha}^{(l)},
\boldsymbol{\pi}^{(l)})$ be an optimal solution to the problem
\eqref{subproblem_rateBl}. If $g(\boldsymbol{\lambda}^{(l)}) \geq
z^{(l)}$, then $(\boldsymbol{\lambda}^{(l)}, z^{(l)})$ is an optimal
solution to (\ref{dualprob_rateBl_re}) (hence
(\ref{dualprob_rateBl})). Otherwise, $(\boldsymbol{\lambda}^{(l)},
z^{(l)})$ is not a solution to the dual problem since it violates the
constraint in (\ref{infCons}) for $\boldsymbol{\alpha} =
\boldsymbol{\alpha}^{(l)}$. In this case, the master problem of next
iteration will be refined by adding a cut, i.e., adding
$\boldsymbol{\alpha}^{(l)}$ to the current collection of points
$\boldsymbol{\alpha}^{(0)},\cdots, \boldsymbol{\alpha}^{(l-1)}$.

Problems (\ref{master_rateBl_dual}) and (\ref{subproblem_rateBl}) are
iteratively solved to find an optimal dual solution
$(\boldsymbol{\lambda}^\star, z^\star)$. The difficulty with huge
dimension has now been encapsulated in problem
(\ref{subproblem_rateBl}) and nicely resolved thanks to the
Proposition \ref{prop1}. The master problem (\ref{master_rateBl_dual})
is a linear program with small dimension (not involving $2^B$ term)
that can be trivially solved using any standard solver. The whole
algorithm is summarized in Algorithm \ref{alg:feasibility}.

\begin{table}[ht]
\caption{Algorithm I: Solving rate balancing problem \eqref{P_rateBalancing} by dual cutting plane}
\label{alg:feasibility}

\begin{boxedalgorithmic}

  \STATE \textbf{Initialization}: Any point $(\boldsymbol{\alpha},
  \boldsymbol{\pi}) \in \mathcal{X}$ can be used as initial point. To
  obtain a sparse solution in particular, we choose
  $(\boldsymbol{\alpha}^{(0)}, \boldsymbol{\pi}^{(0)})$ by activating
  a single pattern randomly and single-BS association for all test
  points. Set $l=0$;

\REPEAT

\STATE $l=l+1$; \STATE Solve \eqref{master_rateBl_dual} by a standard
primal-dual interior-point solver to obtain
$\boldsymbol{\lambda}^{(l)}$ and $z^{(l)}$; \STATE Solve
\eqref{subproblem_rateBl} by Proposition \ref{prop1} to obtain
$(\boldsymbol{\alpha}^{(l)}, \boldsymbol{\pi}^{(l)})$ and
$g(\boldsymbol{\lambda}^{(l)})$;

\UNTIL $g(\boldsymbol{\lambda}^{(l)}) \geq z^{(l)}$;

\STATE Reconstruct an optimal primal solution according to
\eqref{primalRecoverRateBalancin:1} and
\eqref{primalRecoverRateBalancin:2}.
\end{boxedalgorithmic}

\end{table}

The complexity saving of the proposed algorithm in comparison to
standard interior-point solvers can be briefly analyzed as follows. If
problem \eqref{P_rateBalancing} is directly solved by interior-point
methods, the complexity is roughly $\mathcal{O}(I^3K^3B^3)$. By
contrast, every iteration of the proposed algorithm requires finding a
solution to \eqref{subproblem_rateBl} by Proposition \ref{prop1}, and
a solution to \eqref{master_rateBl_dual} by interior-point
solvers. Specifically, solving \eqref{subproblem_rateBl} requires
$\mathcal{O}(IKB)$, while the complexity of solving
\eqref{master_rateBl_dual} depends on the number of constraints in
\eqref{master_rateBl_dual}, which is increased by one inequality per
iteration. Our numerical results suggest that the number of iterations
is proportional to $K$ (see footnote\footnote{The reason that
  Algorithm \ref{alg:feasibility} converges before the number of
  constraints in \eqref{master_rateBl_dual} grows significantly large
  is due to the inherent sparse structure of the solution. As
  identified by Proposition \ref{prop0}, the solution only activates a
  small number of patterns even if all possible patterns are candidate
  ones. Since the proposed algorithm activates one pattern per
  iteration (see \eqref{solu_pi}), the total number of iteration is
  unsurprisingly much lower than $|\mathcal{I}|$ if $|\mathcal{I}|$ is
  large.} for possible reasons). Consequently, it is safe to bound the
complexity of solving \eqref{master_rateBl_dual} as $\mathcal{O}(K^3)$
per iteration. Hence, the overall complexity of the proposed algorithm
is $\mathcal{O}(IK^2B + K^4)$, much lower than directly applying
interior-point solvers to the original problem.

Finally, after the dual problem is solved by the proposed algorithm,
the primal solution can be recovered as follows
\cite[Ch.6]{Bazaraa2013}:
\begin{IEEEeqnarray}{rCl}\label{g}
  \boldsymbol{\alpha}^\star = \sum_{j=0}^{l-1}\kappa_j
  \boldsymbol{\alpha}^{(j)} \label{primalRecoverRateBalancin:1} \\
  \boldsymbol{\pi}^\star = \sum_{j=0}^{l-1}\kappa_j
  \boldsymbol{\pi}^{(j)} \label{primalRecoverRateBalancin:2}
\end{IEEEeqnarray}
where $\kappa_j$ with $j=0,\cdots,l-1$ are the dual variables
associated with the inequality constraints of
\eqref{master_rateBl_dual}, which are typically available as a
by-product if we solve the problem \eqref{master_rateBl_dual} by a
standard interior-point solver.

\subsection{Solving the energy saving problem with rate requirement}
\label{sec:solv-energy-saving}
We now turn the attention to solving (\ref{P1_rateConstrained}) if it
is feasible. One popular approach to handle the $\ell_0$-norm term is
the $\ell_1$-norm approximation. Applying this technique to
(\ref{P1_rateConstrained}), we obtain
\begin{IEEEeqnarray}{rCl}\label{P_rateConst_L1}
  \IEEEyesnumber\IEEEyessubnumber*
  \mathop{\minimize}_{(\boldsymbol{\alpha}, \boldsymbol{\pi}) \in
    \mathcal{X} } \quad && \sum_{b\in\mathcal{B}} P_{b}^{\text{OP}}
  \sum_{k\in\mathcal{K}} \sum_{i \in\mathcal{I}} \alpha_{kbi}    \\
  \textrm{subject to} \quad && \sum_{i \in \mathcal{I}} \sum_{b \in
    \mathcal{B}} \alpha_{kbi} r_{kbi} \geq d_k, \ \forall k
\end{IEEEeqnarray}
where $\mathcal{X}$ is defined by (\ref{const_BS allo}),
(\ref{cons_pi}) and (\ref{con_nonnegative}).

The solutions obtained from (\ref{P_rateConst_L1}) can be further
improved by applying so-called \emph{reweighted $\ell_1$-norm
  minimization methods} \cite{Candes2008}, originally proposed to
enhance the data acquisition in compressed sensing. It is known
that for nonnegative scalar $x \geq 0$, $|x|_0 = \lim_{\epsilon
  \rightarrow 0} \frac{\log(1+x\epsilon^{-1})}{\log(1+\epsilon^{-1})}$
\cite{Pollakis2012}. With a small design parameter $\epsilon >
0$, we neglect the limit and then approximate the $\ell_0$-norm as
\begin{equation}\label{MM_obj}
  |x|_0 \approx \frac{\log(1+x\epsilon^{-1})}{\log(1+\epsilon^{-1})}.
\end{equation}
Relying on (\ref{MM_obj}) and ignoring unnecessary constants, the
problem (\ref{P1_rateConstrained}) can be approximately solved by the
following problem:
\begin{IEEEeqnarray}{rCl}\label{P_approx}
  \IEEEyesnumber\IEEEyessubnumber*
  \mathop{\minimize}_{(\boldsymbol{\alpha}, \boldsymbol{\pi}) \in \mathcal{X}}  \quad &&  \sum_{b\in \mathcal{B}}\left[ (1-q_b) P_b^{\textrm{OP}} \rho_b + \frac{q_b P_b^{\textrm{OP}} \log(\epsilon +\rho_b)}{\log(1+\epsilon^{-1})}  \right] \label{CCP_obj} \IEEEeqnarraynumspace \\
  \textrm{subject to} \quad && \rho_b = \sum_{k\in\mathcal{K}}
  \sum_{i\in\mathcal{I}} \alpha_{kbi}, \  \forall b  \label{CCP_con1} \\
  && \sum_{i \in \mathcal{I}} \sum_{b \in \mathcal{B}} \alpha_{kbi}
  r_{kbi} \geq d_k, \ \forall k. \label{CCP_con2}
 \end{IEEEeqnarray}

Note that (\ref{P_approx}) is a continuous problem unlike the one in
(\ref{P1_rateConstrained}) involving combinatorial terms. However,
(\ref{P_approx}) is not a convex problem since it minimizes a concave
function. Fortunately, it falls into the framework of
difference-of-convex (DC) functions and therefore can be efficiently
solved by the convex-concave procedure \cite{Kuang2012}.

Specifically, by applying the first-order Taylor expansion to the
objective function in (\ref{P_approx}) at the point
$\boldsymbol{\rho}^{(t-1)}$ obtained in $(t-1)$-th iteration, we
arrive at the following problem for the $t$-th iteration:
\begin{IEEEeqnarray}{rCl}\label{P_rateConst_L1_reweighted}
\IEEEyesnumber\IEEEyessubnumber*
  \mathop{\minimize}_{(\boldsymbol{\alpha}, \boldsymbol{\pi}) \in
    \mathcal{X}}  \quad &&
  \sum_{b\in\mathcal{B}} w_b^{(t)} \sum_{k\in\mathcal{K}}\sum_{i\in \mathcal{I}}
  \alpha_{kbi}   \\
  \textrm{subject to}
  \quad 
  &&  \sum_{i \in \mathcal{I}} \sum_{b \in \mathcal{B}}  \alpha_{kbi} r_{kbi} \geq d_k, \ \forall k   \label{consDualrate}
  \end{IEEEeqnarray}
where the weight
\begin{equation}\label{eq_weight}
  w_b^{(t)} = (1-q_b) P_b^{\textrm{OP}} + \frac{q_b P_b^{\textrm{OP}}}{\log(1+\epsilon^{-1}) (\epsilon + \rho_b^{(t-1)} )}
\end{equation}
with
 \begin{equation}\label{eq_cellLoad}
   \rho_b^{(t-1)}= \sum_{k\in\mathcal{K}}
  \sum_{i\in\mathcal{I}} \alpha_{kbi}^{(t-1)}.
 \end{equation}

The convergence can be characterized as follows:
 \begin{prop}\label{prop_convergence}
   Any limiting point of $(\boldsymbol{\alpha}^{(t)},
   \boldsymbol{\pi}^{(t)})$ generated by the above convex-concave
   procedure as $t \rightarrow \infty$ is a stationary point of
   problem \eqref{P_approx}.
\end{prop}

\begin{IEEEproof}
  We first eliminate the equality constraint \eqref{CCP_con1} by
  substituting it into the objective \eqref{CCP_obj}, and then denote
  the rest of constraint set ($\mathcal{X}$ with \eqref{CCP_con2}) by
  $\mathcal{Y}$. It can be easily verified that $\mathcal{Y}$ is
  compact (closed and bounded). According to Remark 7 in
  \cite{Lanckriet2009}, our problem satisfies all conditions of
  Theorem 4 in \cite{Lanckriet2009}. By applying this theorem, the
  proposition is proved.
\end{IEEEproof}
 In practice, the reweighted $\ell_1$ method converges typically within 6-10
 iterations to a desirable accuracy and the largest improvement in sparsity is obtained in the
 first few iterations.

 Problem \eqref{P_rateConst_L1_reweighted} is a linear program. Like
 in Section \ref{sectionFeasiTest}, it can be tackled in the dual
 domain by the cutting plane method, resulting in similar complexity
 saving as explained in Section \ref{sectionFeasiTest}. In Section
 \ref{sec:performanceProposed}, we will compare the running time of
 the proposed algorithm to that of a commercial solver by simulation.

 By dualizing the constraint of (\ref{consDualrate}), we
can express the dual function as
\begin{equation}\label{P_rateConst_Dualfun}
  h(\boldsymbol{\mu}) = \inf_{(\boldsymbol{\alpha}, \boldsymbol{\pi})
    \in \mathcal{X}} \{  \sum_{k,b,i}
    \alpha_{kbi} w_b^{(t)} -\sum_{k,b,i} \alpha_{kbi}r_{kbi} \mu_k   +\sum_k d_k \mu_k   \}
\end{equation}
where  $\boldsymbol{\mu} =
(\mu_1,\cdots, \mu_K)^T$ is the Lagrangian multiplier.

Following the idea presented in Section \ref{sectionFeasiTest}, we
formulate the master problem as
\begin{IEEEeqnarray}{rCl}\label{dualprob_energy_master}
  \IEEEyesnumber
  \mathop{\maximize}_{\boldsymbol{\mu}\geq 0, z}  \ \ &&  z \IEEEyessubnumber \\
  \textrm{subject to} \   && \sum_{k,b,i}\alpha_{kbi}^{(t,j)} w_b^{(t)} -\sum_{k,b,i}  \alpha_{kbi}^{(t,j)} r_{kbi} \mu_k   +\sum_k d_k \mu_k  \geq z, \IEEEnonumber\\
   && \qquad \qquad \qquad \qquad \qquad \forall j\in \{0,\cdots, l-1\} \IEEEyessubnumber \label{consmaster}
\end{IEEEeqnarray}
and the inner problem as
\begin{equation}\label{dualprob_energy_inner}
  \mathop{\minimize}_{(\boldsymbol{\alpha}, \boldsymbol{\pi}) \in
    \mathcal{X}} \quad \sum_{k,b,i} \alpha_{kbi}  \left(w_b^{(t)} -  r_{kbi}  \mu_k^{(l)} \right)  +\sum_k d_k \mu_k^{(l)}
\end{equation}
respectively, where we denote the solution to
(\ref{dualprob_energy_master}) by $(\boldsymbol{\mu}^{(l)}, z^{(l)})$
and the solution to (\ref{dualprob_energy_inner}) by
$(\boldsymbol{\alpha}^{(t,l)}, \boldsymbol{\pi}^{(t,l)})$. The master
problem (\ref{dualprob_energy_master}) is refined for the next
iteration by adding $\boldsymbol{\alpha}^{(t,l)}$ to the constraint (\ref{consmaster}). In
this way, we iteratively solve (\ref{dualprob_energy_master}) and
(\ref{dualprob_energy_inner}) until $h(\boldsymbol{\mu}^{(l)}) \geq
z^{(l)}$, implying that we have solved the problem
(\ref{P_rateConst_L1_reweighted}) in the dual domain. Then we can
find the primal solution following the same idea of
\eqref{primalRecoverRateBalancin:1} and
\eqref{primalRecoverRateBalancin:2} as:
\begin{IEEEeqnarray}{rCl}
  \boldsymbol{\alpha}^{(t)} = \sum_{j=0}^{l-1}\kappa_j
  \boldsymbol{\alpha}^{(t,j)} \label{eq:1} \\
  \boldsymbol{\pi}^{(t)} = \sum_{j=0}^{l-1}\kappa_j
  \boldsymbol{\pi}^{(t,j)} \label{eq:2}
\end{IEEEeqnarray}
where $\kappa_j$ with $j=0,\cdots,l-1$ are the dual variables
corresponding to the inequality constraints of \eqref{consmaster},
which are available if we solve the master problem
\eqref{dualprob_energy_master} by off-the-shelf interior-point solvers.

Finally, the outermost iteration is to adjust the weights according to
(\ref{eq_weight}) and (\ref{eq_cellLoad}) and then the problem
(\ref{P_rateConst_L1_reweighted}) is solved again with the new
weights. We summarize the proposed approach in Algorithm \ref{alg1:solv-energy-saving}.

\begin{table}[ht]
\caption{Algorithm II: Energy saving with the user rate constraint}
\label{alg1:solv-energy-saving}
\begin{boxedalgorithmic}
\STATE  \textbf{Feasibility check}: Solve the rate balancing problem
  \eqref{P_rateBalancing} by the methods described in Section
  \ref{sectionFeasiTest} to obtain
  $\boldsymbol{\alpha}^\star_\textrm{bln},\boldsymbol{\pi}^\star_\textrm{bln}, \textrm{and}, R_{\textrm{sum}}^\star$. If
   $R_{\textrm{sum}}^\star  < \sum_k d_k$, then
  the user rate constraint is infeasible; Else if
  $R_{\textrm{sum}}^\star  = \sum_k d_k$, problem is solved; Otherwise we proceed to the
  next step;

  \STATE \textbf{Initialization}: Outer iteration counter $t=0$,
  $\boldsymbol{\alpha}^{(0)}=\boldsymbol{\alpha}^\star_\textrm{bln}$,
  $\boldsymbol{\pi}^{(0)}=\boldsymbol{\pi}^\star_\textrm{bln}$;
  \REPEAT \STATE $t=t+1$; \STATE Update the weights: If $t=1$, then
  $\{w_b^{(1)}= P_b^{\text{OP}}, \forall b\}$ as given in \eqref{P_rateConst_L1}; otherwise
  calculate $\{w_b^{(t)},\forall b\}$ according to \eqref{eq_weight} and
  \eqref{eq_cellLoad};

\STATE Initialize inner iteration counter $l=0$,  $\boldsymbol{\alpha}^{(t,0)}=\boldsymbol{\alpha}^\star_\textrm{bln} $,
$\boldsymbol{\pi}^{(t,0)}= \boldsymbol{\pi}^\star_\textrm{bln}$;
\REPEAT

\STATE $l = l+1$;
\STATE Solve \eqref{dualprob_energy_master} by a standard primal-dual
interior-point solver to obtain
$\boldsymbol{\mu}^{(l)}$, $z^{(l)}$;
\STATE Solve \eqref{dualprob_energy_inner} by Proposition \ref{prop2} to obtain
$\boldsymbol{\alpha}^{(t,l)}$, $\boldsymbol{\pi}^{(t,l)}$ and
$h(\boldsymbol{\mu}^{(l)})$;

\UNTIL $h(\boldsymbol{\mu}^{(l)}) \geq z^{(l)}$;
\STATE Reconstruct an optimal primal solution
$(\boldsymbol{\alpha}^{(t)}, \boldsymbol{\pi}^{(t)} ) $ from
$( \boldsymbol{\alpha}^{(t,j)}, \boldsymbol{\pi}^{(t,j)}),  j=0,\cdots,l-1$
  according to \eqref{eq:1} and \eqref{eq:2};

\UNTIL Objective \eqref{CCP_obj} converges
or the maximum number of iterations is reached.
\end{boxedalgorithmic}
\end{table}

The key enabler in Algorithm \ref{alg1:solv-energy-saving} is the following Proposition
\ref{prop2}, which shares the similar spirit to Proposition
\ref{prop1}:

\begin{prop}\label{prop2}
  The problem \eqref{dualprob_energy_inner} has a closed-form solution
  that can be expressed as
  \begin{equation}\label{eq:3}
    \alpha_{kbi}^{(t,l)} =    \left\{  \begin{array}{rl}
        1 & \text{if} \ i=\bar{i}, k=\bar{k}(b, \bar{i}), \
        \text{and}\  \tilde{r}_{kbi} < 0  \\
        0 & \text{otherwise}
      \end{array} \right.
  \end{equation}
  and
  \begin{equation}\label{eq:4}
    \pi_i^{(t,l)} =    \left\{  \begin{array}{rl}
        1 & \text{if} \ i=\bar{i}  \\
        0 & \text{otherwise}
      \end{array} \right.
  \end{equation}
where $\bar{k}(b,i)=\arg \min_k \tilde{r}_{kbi}$ with $\tilde{r}_{kbi}
=  w_b^{(t)} - r_{kbi}\mu_k^{(l)}$, and $\bar{i} =
\arg \min_i  \sum_b\left[
  \tilde{r}_{\bar{k}(b,i)bi}\right]^- $, where $[x]^- =\min(0,x)$.
\end{prop}

\begin{IEEEproof}
  The proof can be obtained following the similar steps that we use to prove Proposition
  \ref{prop1}. The only difference is that $\tilde{r}_{kbi}$ can now take
  positive values. If $\min_k \tilde{r}_{kbi} >0$ for the given $b$ and $i$,
  then
  \begin{IEEEeqnarray}{rCl}
    \mathop{\minimize}_{\alpha_{kbi}\geq 0, \sum_k
    \alpha_{kbi}\leq \pi_i} \quad  \sum_{k,b,i}\alpha_{kbi} \tilde{r}_{kbi}
  \end{IEEEeqnarray}
will result in $\alpha_{kbi} = 0, \forall k$ for the given $b$ and
$i$. Moreover, if $(\min_k \tilde{r}_{kbi}) =0$ for the given $b$ and
$i$, we can also set $\alpha_{kbi}=0$  without affecting the optimality.
 Taking these two facts into account, we modify $\alpha_{kbi}^{(t,l)}$ in \eqref{eq:3} accordingly,
and $\bar{i}$ as well.
\end{IEEEproof}

Several remarks to the Algorithm \ref{alg1:solv-energy-saving} are as follows.
\begin{myremark}
  The cutting plane method should be initialized with a strictly
  primal feasible solution, otherwise the master problem will become
  unbounded in the first iteration or not work properly. The proposed
  Algorithm \ref{alg1:solv-energy-saving} checks the feasibility of the primal problem in step 1. If the prescribed rate
  constraints are feasible, the checking procedure guarantees to find
  a strictly primal feasible solution, which is used to initialize the
  cutting plane iteration (see step 6).
\end{myremark}

\begin{myremark}
  In step 5, this particular choice of the initial weighting matrix
  means that we solve the (unweighted) $\ell_1$-norm approximation problem
  \eqref{P_rateConst_L1} directly in the first iteration.
\end{myremark}

\section{Unified study of baseline strategies}
\label{sec:unif-study-baseline}

In this section, we show how to develop a unified view on a wide range
of previous strategies. With this view, we can analyze and compare
various resource management strategies in a unified framework.

\subsection{Cell activation and user association without interference
  coordination}
\label{sec:noICIC}

The cell activation and user association has been studied in
\cite{Pollakis2012,Cavalcante2014}, where worst-case estimates of the
user rates resulted from no intercell interference coordination are
used to calculate the QoS requirements.  The strategy presented in
\cite{Pollakis2012,Cavalcante2014} can be easily analyzed using the
proposed framework in this paper. It corresponds to restricting the
candidate pattern set to exactly one pattern: All-ON pattern (i.e.,
reuse-1 pattern).  To compute the link rate under the reuse-1
pattern, we set $\{a_{il}=1, \forall l\in \mathcal{B} \}$ in
\eqref{SINR} and then calculate the rate according to
\eqref{RatePerCarrier}. Since there is only one allowed pattern for
resource allocation, we can drop both the subscript $i$ and the pattern
allocation variable $\boldsymbol{\pi}$, focusing on the two-dimension
resource allocation variable $\alpha_{kb}$ only.

Consequently, the user rate constrained energy saving problem of
\eqref{P1_rateConstrained} boils down to
\begin{IEEEeqnarray}{rCl}
  \label{eq:21}\IEEEyesnumber\IEEEyessubnumber*
\mathop{\minimize}_{\{\alpha_{kb}\}}
  \quad && P^{\text{tot}}= \sum_{b\in \mathcal{B}} \left[ (1-q_b)\rho_b P_b^\textrm{OP} +
    q_b |\rho_b|_0 P_b^\textrm{OP} \right] \IEEEeqnarraynumspace \\
\text{subject to}  \quad  && \rho_b = \sum_{k\in \mathcal{K}}
\alpha_{kb}, \forall b \\
     && \sum_{b\in\mathcal{B}} \alpha_{kb} r_{kb} \geq d_k, \forall k
     \\
    &&  \sum_{k \in \mathcal{K}} \alpha_{kb} \leq 1, \forall b \\
&&    \alpha_{kb} \geq 0, \forall k,b
\end{IEEEeqnarray}
which can be solved by the convex-concave procedure as described
before. We will compare this strategy with our proposal in Section \ref{sec:results}.


\subsection{Range expansion user association}
\label{sec:range-expansion-user}
Reference signal received power (RSRP) is adopted in LTE/LTE-A
standards as a signal quality indicator \cite{3GPP214}.
In LTE-A HetNets, range expansion (RE) has been further introduced as a simple
scheme to control the load distribution among pico and macro layers
\cite{Madan2010}. The basic mechanism is to add a positive bias (in
dB) to the RSRP received from small cells when deciding the association of user
equipments (UEs).

In the proposed framework, to express the association of the test
point $k$ according the RE rule, a binary association indicator
$s_{kb}$ can be introduced:
  \begin{equation}\label{RE_asso}
    s_{kb} =    \left\{  \begin{array}{rl}
        1 & \text{if} \ b=\arg\max_{j\in \mathcal{B}}\left(\text{RSRP}_{jk}+\eta_j \right)
          \\
        0 & \text{otherwise}
      \end{array} \right., \  \quad \forall k
  \end{equation}
where $\text{RSRP}_{jk}$ is the received RSRP (in dBm) at the test
point $k$ from cell $j$, and $\eta_j$ is the bias value (in dB) of cell $j$.

\subsubsection{Cell activation and interference coordination with
  fixed RE association }
\label{sec:cell-activ-interf}

In order to separate the impact of the RE user association rule from
that of resource allocation, we can pre-define a set of RE biases
and fix the user association according to \eqref{RE_asso}. Then the
cell activation and resource allocation problem with fixed user
association can be formulated as
\begin{IEEEeqnarray}{rCl}\label{eq:5}
\IEEEyesnumber\IEEEyessubnumber*
  \displaystyle\mathop{\minimize}_{\boldsymbol{\alpha},\boldsymbol{\pi}}
  \quad
  &&    P^{\text{tot}}= \sum_{b\in \mathcal{B}} \left[ (1-q_b)\rho_b P_b^\textrm{OP} +
    q_b |\rho_b|_0 P_b^\textrm{OP} \right]  \label{eq:6} \IEEEeqnarraynumspace\\
  \text{subject to} \quad && \rho_b =  \sum_{k\in\mathcal{K}_b}
  \sum_{i\in\mathcal{I}} \alpha_{kbi}, \  \forall b \label{eq:7} \\
  && \sum_{i \in \mathcal{I}} \sum_{b \in \mathcal{B}}  \alpha_{kbi} r^\text{RE}_{kbi} \geq d_k, \ \forall k  \label{eq:8} \\
  && \sum_{k \in \mathcal{K}_b} \alpha_{kbi} \leq \pi_i, \forall b,\ \forall i \label{eq:9}\\
  && \sum_{i \in \mathcal{I}} \pi_i = 1 \label{eq:10}  \\
  && \pi_i \geq 0, \ \forall i, \quad \alpha_{kbi} \geq 0, \ \forall
  k,b,i   \label{eq:11}
\end{IEEEeqnarray}
where $r_{kbi}^\text{RE} \triangleq s_{kb}r_{kbi}$ is the effective
rate obtained by forcing the elements of $r_{kbi}$ that are not
allowed to associate due to the RE rule to zero, and
$\mathcal{K}_b = \{k\in\mathcal{K} : s_{kb} =1\}$.

The above formulation is almost the same as the problem
\eqref{P1_rateConstrained} with the only two differences. First, the
rate $r_{kbi}$ in problem \eqref{P1_rateConstrained} is replaced by
the effective rate $r^\text{RE}_{kbi}$. Second, the summations in
\eqref{eq:7} and \eqref{eq:9} are over $\mathcal{K}_b$ instead of
$\mathcal{K}$. Interestingly, the following Proposition \ref{prop_eq}
justifies that $\mathcal{K}_b$ can be replaced by $\mathcal{K}$
without loss of optimality, meaning that the algorithms developed in
Section \ref{sec:rate-constr-energy} can be directly applied to solve
\eqref{eq:5}.

\begin{prop}\label{prop_eq}
  The problem \eqref{eq:5} can be equivalently solved by replacing all
  $\mathcal{K}_b$ with $\mathcal{K}$.
\end{prop}

\begin{IEEEproof}
  By replacing $\mathcal{K}_b$ with $\mathcal{K}$ in problem
  \eqref{eq:5}, we restrict the feasible set since $\mathcal{K}_b
  \subseteq \mathcal{K}$. However, doing so will not compromise the
  optimality. The reason is that BS $b$ does not contribute any rate
  for users outside $\mathcal{K}_b$ (due to the definition of
  $r^\text{RE}_{kbi}$). Hence, we can set $\alpha_{kbi}=0$, if
  $k\notin \mathcal{K}_b$. Formally, the proof is given as follows.

  In this proof, we refer to the new problem where $k \in
  \mathcal{K}_b$ in \eqref{eq:7} and \eqref{eq:9} has been replaced
  with $k \in \mathcal{K}$ as the \emph{reformulated problem}.  Let
  $(\boldsymbol{\alpha}^\text{new}$, $\boldsymbol{\pi}^\text{new})$ be
  the solution to this reformulated problem. Accordingly in the
  reformulated problem, define $\rho_b^\text{new} = \sum_{k\in
    \mathcal{K}} \sum_{i \in \mathcal{I}} \alpha^\text{new}_{kbi}$,
  $\forall b\in \mathcal{B}$. Note that
  $(\boldsymbol{\alpha}^\text{new}$, $\boldsymbol{\pi}^\text{new})$
  satisfies the original constraints from \eqref{eq:8} to
  \eqref{eq:11}, since $r_{kbi}^\text{RE} \geq 0$, $\mathcal{K}_b
  \subseteq \mathcal{K}$ and $\alpha_{kbi} \geq 0$. Next we prove that
  $(\boldsymbol{\alpha}^\text{new}$, $\boldsymbol{\pi}^\text{new})$
  must be the solution to problem \eqref{eq:5} by contradiction.

Suppose this is not true, meaning that we can find
  another feasible point $(\boldsymbol{\alpha}^\text{old}$,
  $\boldsymbol{\pi}^\text{old})$ in problem \eqref{eq:5} such that
  $P^\text{tot}(\boldsymbol{\rho}^\text{old}) <
  P^\text{tot}(\boldsymbol{\rho}^\text{trim})$, where
  $\boldsymbol{\rho}^\text{old} =
  (\rho_1^\text{old},\cdots,\rho_B^\text{old})^T$ with
  $\rho_b^\text{old} = \sum_{k\in \mathcal{K}_b} \sum_{i \in
    \mathcal{I}} \alpha^\text{old}_{kbi}$, and $\boldsymbol{\rho}^\text{trim} =
  (\rho_1^\text{trim},\cdots,\rho_B^\text{trim})^T$ with
  $\rho_b^\text{trim} = \sum_{k\in \mathcal{K}_b} \sum_{i \in
    \mathcal{I}} \alpha^\text{new}_{kbi}$.

 In such case, we can
  construct another point $(\boldsymbol\alpha^\prime,
  \boldsymbol{\pi}^\prime)$ by choosing $\alpha_{kbi}^\prime =
  \left\{ \begin{array}{rl}
      0 & \text{if} \ k \notin \mathcal{K}_b  \\
      \alpha_{kbi}^\text{old} & \text{otherwise}
    \end{array} \right.$  and
  $\boldsymbol\pi^\prime =
  \boldsymbol\pi^\text{old}$,
  respectively. It can be easily seen that $(\boldsymbol\alpha^\prime,
  \boldsymbol{\pi}^\prime)$ is also feasible in the reformulated problem. Define  $\rho_b^\prime = \sum_{k\in \mathcal{K}} \sum_{i \in
    \mathcal{I}} \alpha^\prime_{kbi}$, $\forall b$. We arrive at
  $P^\text{tot}(\boldsymbol{\rho}^\prime) =
  P^\text{tot}(\boldsymbol{\rho}^\text{old}) <
  P^\text{tot}(\boldsymbol{\rho}^\text{trim}) \leq
  P^\text{tot}(\boldsymbol{\rho}^\text{new}) $. In other words, we
  find a feasible point $(\boldsymbol\alpha^\prime,
  \boldsymbol{\pi}^\prime)$ in the reformulated problem that gives
  lower value of the objective function than $(\boldsymbol{\alpha}^\text{new}$,
  $\boldsymbol{\pi}^\text{new})$, which is contradictory to the
  optimality of $(\boldsymbol{\alpha}^\text{new}$,
  $\boldsymbol{\pi}^\text{new})$ in the reformulated problem.
\end{IEEEproof}

\subsubsection{Mapping the jointly optimized association to cell-specific
  biases}
\label{sec:mapp-optim-assoc}

In Section \ref{sec:rate-constr-energy}, we propose and solve the
coupled problem of optimizing the user association, cell activation
and resource allocation. It is interesting to see how this jointly
optimized user association can be mapped to the cell-specific
biases. In other words, we would like to choose values of $\eta_j,
\forall j\in \mathcal{B}$ such that the user association based on the
rule given by \eqref{RE_asso} leads to the association decisions
derived in Section \ref{sec:rate-constr-energy} by a joint
optimization.

To achieve this goal, we propose to minimize a weighted mean square
error of the association. Specifically, we aim to solve the following
optimization problem:
\begin{IEEEeqnarray}{rCl}\label{eq:12}
  \boldsymbol{\eta}^\star = \arg \min_{ \boldsymbol{\eta} \geq 0 }
  \quad \sum_{b \in \mathcal{B}} \omega_b \sum_{k\in \mathcal{K}}
  \left( s_{kb}(\boldsymbol{\eta}) - s_{kb}^\star \right)^2
\end{IEEEeqnarray}
where $\boldsymbol{\eta} = (\eta_{1}, \cdots,
\eta_{B})^T$, the weight $\omega_b$ is used to
emphasize the different impacts of the association error on different
cells (e.g., macro cells can have larger weights to account for the
larger energy consumption if macro cells are forced to switch on due
to errors in the user association), $s_{kb}$ is related to
$\boldsymbol{\eta}$ according to \eqref{RE_asso}, the reference
association $s^\star_{kb}$ is derived from the joint optimization
given in Section \ref{sec:rate-constr-energy} as $s^\star_{kb} =
\left| \sum_{i\in \mathcal{I}} \alpha_{kbi}^\star\right|_{0}$,
$\forall k, \forall b$, and $\alpha_{kbi}^\star$ is
the solution to problem \eqref{P1_rateConstrained} using the Algorithm
\ref{alg1:solv-energy-saving}.

Problem \eqref{eq:12} is solved by a coordinate descent method in this
paper, i.e., one-dimensional search for one bias value is performed at
a time while keeping the rest of biases fixed. The iterative procedure
is terminated if the objective cannot be further reduced. The
objective function is guaranteed to converge, although the resulting solution
is not necessarily globally optimal. The solution depends on the order
of BSs being updated. Hence, we solve \eqref{eq:12} several times with
different updating orders, and then select the best one. Also note
that the RE rule typically results in the single-BS association for
each test points, but the solution derived by joint optimization in
Section \ref{sec:rate-constr-energy} may yield multiple-BS association
for some test points. Hence, in general the minimum error in
\eqref{eq:12} is greater than zero. Nevertheless, as will be shown in
Section \ref{sec:simuBias}, the proposed method for solving
\eqref{eq:12} gives a set of nearly optimal cell-specific biases.

\subsection{Cell clustering for activation }
\label{sec:cell-clust-activ}

In order to reduce the number of candidate patterns for resource
allocation, we can group multiple cells into clusters. Then each
cluster is regarded as one giant cell when formulating interference
patterns. In other words, all cells within the same cluster can only
be simultaneously activated or deactivated.

The principle of forming a cluster is to group cells that do not
interfere with each other or have very weak mutual interference into
one cluster.  In this way, simultaneously activating them will not
significantly increase the network interference. In HetNets, pico BSs
have low transmit power and antenna gain.  So the interference among
pico cells is expected to be low if they are deployed with reasonable
inter-site distances. Therefore, we can group pico cells within one
macro cell into one cluster.  One the other hand, we separate the
dominant mutually interfering cells (e.g., pico cells and its umbrella
macro cell) into different clusters, such that the inter-cluster
interference can be handled by resource allocation among clusters.

The framework proposed in Section \ref{sec:rate-constr-energy} can be
used to evaluate various cell clustering strategies from the energy
saving perspective. Cell clustering results in a restricted set of
candidate patterns. The user association and resource allocation can
then be performed over this pattern set. The algorithms developed in
Section \ref{sec:rate-constr-energy} can be directly applied.


\section{Performance Evaluation}
\label{sec:results}

\subsection{Simulation setup}

\begin{figure}[!t]
\centering
\includegraphics[width=3.2In]{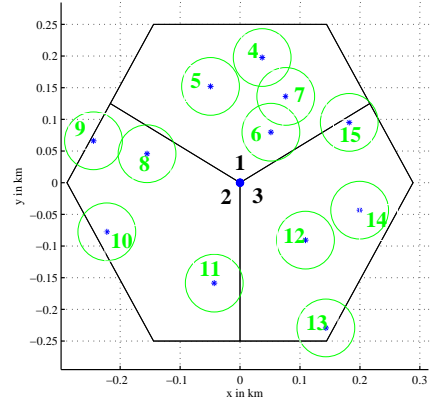}
\caption{A heterogeneous network consisting of 15 cells. }\label{fig_network}
\end{figure}

A network consisting 15 cells have been used in the simulations. Among
these cells, there are 3 macro cells, each of which contains 4
randomly dropped pico cells, as shown in Fig.~\ref{fig_network}. The
cells are labeled as
\begin{equation*}
  \underbrace{1,2,3}_{\text{macro cells}}, \underbrace{4,5,6,7}_\text{picos in cell 1}, \underbrace{8,9,10,11}_\text{picos in cell 2}, \underbrace{12,13,14,15}_\text{picos in cell 3}.
\end{equation*}

The parameters for propagation modelling follow the suggestions in
3GPP evaluation methodology \cite{3GPP2010}, and are summarized in
Table \ref{tab:1} together with other system parameters. Based on the
linear relationship between transmit power and operational power
consumption\footnote{We adopted the linear model in \cite{Fehske2009}:
  $P_b^\text{OP} =\alpha_b P_b +\beta_b $, where $P_b$ is the transmit
  power for BS $b$, $\alpha_b= \frac{22.6}{3}$ and $\beta_b =
  \frac{412.4}{3} $W if $b$ is a macro; otherwise $\alpha_b = 5.5$ and
  $\beta_b = 32 $W if $b$ is a pico.}, we calculate the maximum
operational power $P^\text{OP}$ as 439W and 38W for macro and pico
BSs, respectively. We further assume each macro BS has a constant
power consumption, i.e., $q_b =1$, $\forall b \in
\mathcal{B}_\text{macro}$, and the fixed power consumption of a pico
takes $50\%$ of the maximum operational power, i.e. $q_b = 0.5$,
$\forall b \in \mathcal{B}_\text{pico}$. Note that these assumptions
are made for providing concrete numerical results, and they are not
from the restriction of our formulation.  For reweighted $\ell_1$-norm
minimization, we set $\epsilon = 10^{-6}$ and the maximum number of
iteration is 10.

\begin{table}[!t]
\renewcommand{\arraystretch}{1.0}
\caption{Network parameters.}
\label{tab:1}
\centering
\begin{tabular}{c c}
\hline\hline 
Parameter & Description \\ [0.5ex] 
\hline 
bandwidth & 10 MHz  \\ 
Macro total Tx power   & 46 dBm  \\
Macro $P^\text{OP}$ and $q_b$ & 439 W, 1 \\
Pico total Tx power    & 30 dBm \\
Pico $P^\text{OP}$ and $q_b$& 38 W, 0.5 \\
Macro antenna gain & 15 dB\\
Pico antenna gain & 5 dB \\
Macro path loss & $128.1+37.6\log_{10}(R)$ \\
Pico path loss & $140.7+36.7 \log_{10}(R)$ \\
Penetration loss & 20 dB \\
Shadowing std. dev. & 8dB(macro), 10dB(pico) \\
Shadowing corr. distance & 25 m \\
Macrocell shadowing corr. & 1 between cells \\
Picocell shadowing corr. & 0.5 between cells \\
Fading model & No fast fading \\
Min. macro(pico)-UE dist. & 35 m (10 m) \\
Min. macro(pico)-pico dist. & 75 m (40 m)\\
Noise density and noise figure & -174 dBm/Hz, 9dB\\
 \hline\hline 
\end{tabular}
\end{table}

\subsection{Performance of the proposed algorithm}
\label{sec:performanceProposed}

\begin{figure}

\centering
\begin{minipage}{.45\textwidth}
\centering
\includegraphics[width=\linewidth]{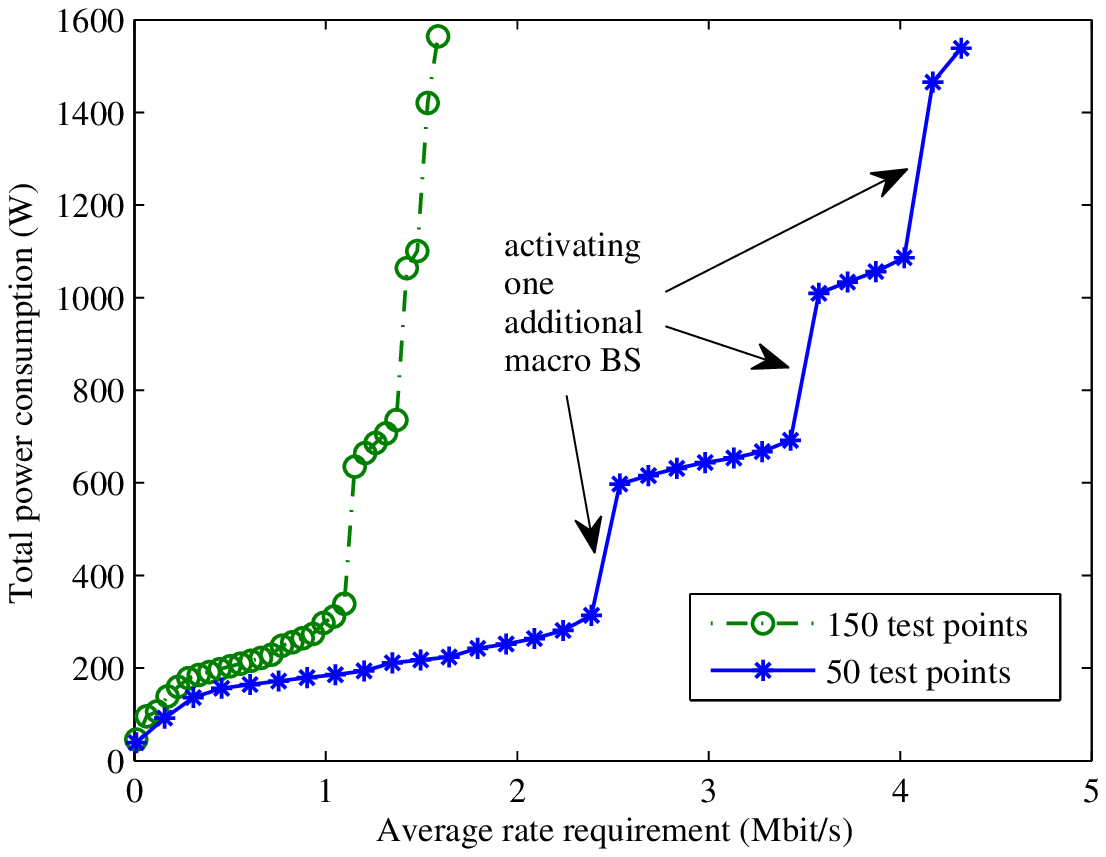}
\caption{Network power consumption achieved by the proposed Algorithm \ref{alg1:solv-energy-saving}, where all test points are assumed to have the same rate requirement.}
\label{fig:result1}
\end{minipage}\hfill
\begin{minipage}{.45\textwidth}
\centering
\includegraphics[width=\linewidth]{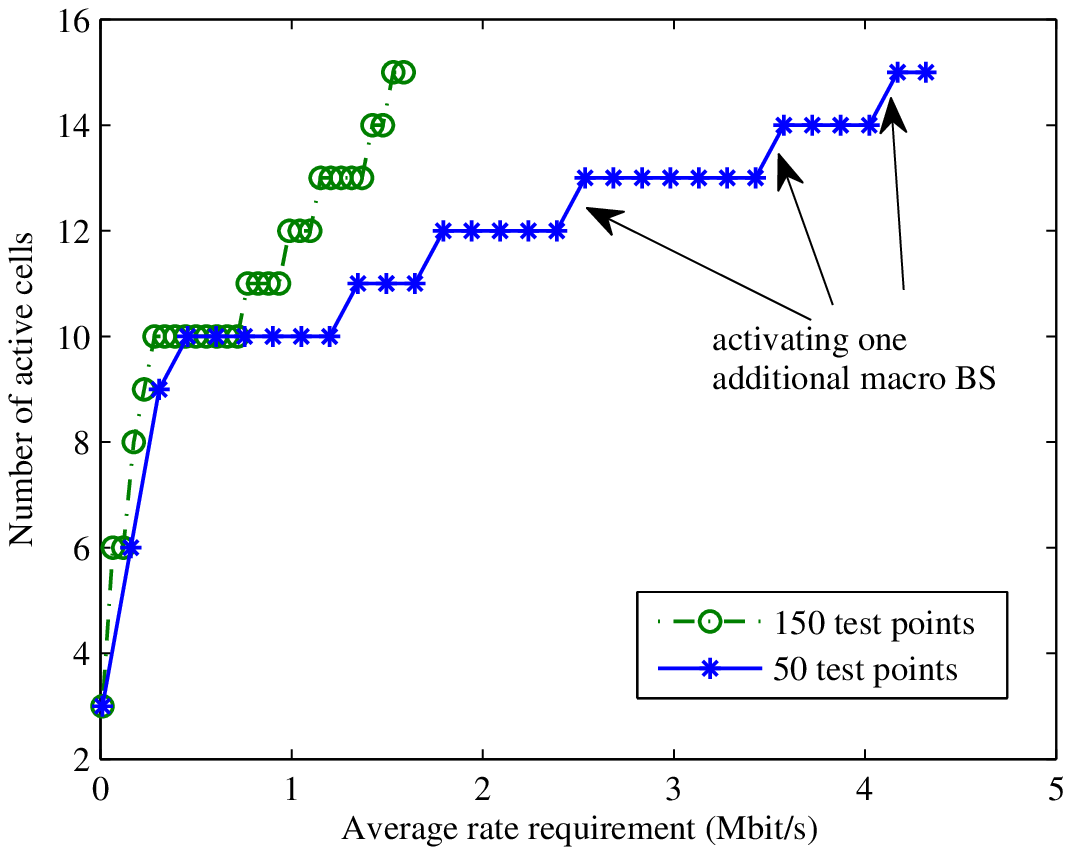}
\caption{Number of active cells achieved by the proposed Algorithm \ref{alg1:solv-energy-saving}, where all test points are assumed to have the same rate requirement.}
\label{fig:result2}
\end{minipage}\hfill

\end{figure}

Fig.~\ref{fig:result1} and Fig.~\ref{fig:result2} report the network
power consumption and number of active BSs, respectively, obtained by
the proposed Algorithm \ref{alg1:solv-energy-saving}. The results are plotted versus the rate
requirement of the test points, where 50 and 150 test points are
uniformly distributed within the network after dropping the picos, and
all test points are assumed to have the same rate requirement for
simplicity.

As shown in Figs.~\ref{fig:result1} and \ref{fig:result2}, the network
power consumption, as well as the number of active BSs, increases with
the user rate requirement. The maximum feasible rate requirement for
50 test points in the considered network realization is $d_k =
4.32$Mbit/s, $\forall k$, while it reduces to $1.58$Mbit/s if 150 test
points need to be supported. In Figs. \ref{fig:result1} and
\ref{fig:result2}, the rate requirements are chosen uniformly between
$0.01$Mbit/s and the maximum feasible rates. Interestingly, we observe
3 steep increases in the power consumption in Fig.~\ref{fig:result1}
for both 50-test-point-case and 150-test-point-case. Actually, each of
these jumps corresponds to activating one macro BS. Note that the
proposed algorithm successfully deactivates the macro BSs for power
saving when the rate requirement is small or moderate. It activates
macro BSs only if necessary.

 \begin{table}[t]
\renewcommand{\arraystretch}{1.0}
\caption{Number of active patterns after Algorithm \ref{alg1:solv-energy-saving} converges. }
\label{tab:numActivePattern}
\centering
\begin{tabular}{|c | c | c | c |c | c| }
\hline
Rate requirement (Mbit/s) &  0.1  &  0.5   &  1.0  &  1.5  &  2.0 \\
\hline
\hline
50 test points & 5   & 9 & 18 & 18 & 23 \\
\hline
150 test points  &  6 & 25  & 43 & 47  & infeasible \\
\hline 
\end{tabular}
\end{table}

To verify the sparsity structure of the solution indicated by
Proposition \ref{prop0}, we show in Table~\ref{tab:numActivePattern}
the number of active patterns after Algorithm \ref{alg1:solv-energy-saving} converges when all
possible $2^{15}$ patterns are considered as the candidates. As shown,
the solution found by the proposed algorithm indeed allocates
resources to a small number of patterns. Most of the candidate
patterns have not been used. The number of active patterns slightly
increases with the rate requirement, but is less than the bound established in Proposition \ref{prop0}.

We further show the average number of serving BSs for each test point
in Table~\ref{tab:numMultiAssociation}. As shown, although problem
formulation of \eqref{P1_rateConstrained} does not enforce single-BS
association constraints, almost all test points are associated with
single BS as the result of optimization. In the simulated cases,
multiple-associated test points only occur in three scenarios: 50 test
points with demands of $2.0$Mbit/s, 150 test points with $1.0$Mbit/s
and $1.5$Mbit/s. In these three scenarios, only 1 out of 50, 4 out of
150 and 3 out of 150 are multiple-associated test points,
respectively. The number of serving BSs for these multiple-associated
test points are two.

To enforce single association for ALL test points, we can introduce
the binary association indicator $s_{kb}$ and apply range expansion
according to \eqref{RE_asso}, where the cell-specific biases can be
obtained from the optimal user resource parameter $\alpha_{kbi}$, as
given by \eqref{eq:12}. This range expansion based single association
will facilitate the implementation in practice. The performance will
be evaluated in the following sections.

 \begin{table}[t]
\renewcommand{\arraystretch}{1.0}
\caption{Average number of serving BSs per test point. }
\label{tab:numMultiAssociation}
\centering
\begin{tabular}{|c | c | c | c |c |c| }
\hline
Rate requirement (Mbit/s) &  $0.1$  &  $0.5$   &  $1.0$  &  $1.5$  & $2.0$ \\
\hline
\hline
50 test points & 1   & 1 & 1 & 1 & 1.020  \\
\hline
150 test points  &  1 & 1  & 1.027 & 1.020 & infeasible \\
\hline 
\end{tabular}
\end{table}

 \begin{table}[t]
\renewcommand{\arraystretch}{1.0}
\caption{Algorithm running time. }
\label{tab:runTime}
\centering
\begin{tabular}{|c | c | c | c |c | }
\hline
Number of candidate patterns &  $19$  &  $2^6$   &  $2^9$  &  $2^{15}$  \\
\hline
\hline
Proposed algorithm (sec) & 4.2   & 10.2 & 13.8 & 313  \\
\hline
Gurobi solver (sec)  &  0.3 & 1.2  & 19.6 & 6346 \\
\hline 
\end{tabular}
\end{table}


Finally, we compare the running time of Algorithm \ref{alg1:solv-energy-saving}
to that of replacing steps 6 to 12 with state-of-the-art commercial solver, Gurobi \cite{gurobi} (with barrier method selected), and report the
results in Table~\ref{tab:runTime}. In the simulated case, $50$ test
points are uniformly distributed in the network with the same demand
of $200$kbit/s. The results are averaged over $10$ random drops. The
algorithms are executed in Matlab $2014$ on an Intel Core i$7$
$2.2$GHz quad-core computer with $8$GB RAM. We apply clustering to
obtain different number of candidate patterns
as shown in Table~\ref{tab:runTime} (see Section
\ref{sec:comparingMethods} for details of feature pattern selection).  Compared
to the industrial-strength solver, our algorithm
with a unsophisticated implementation starts to achieve some gains as
the problem dimension grows. In particular, when $2^{15}$ patterns are
considered, a significant saving in running time is observed. To conclude, the proposed algorithm provides a feasible
way to calculate the benchmark considering all $2^B$ patterns in a
reasonable-sized network. It can also be applied to larger networks
over a set of preselected patterns by clustering, still achieving
complexity saving.

\subsection{Comparing different strategies}
\label{sec:comparingMethods}

In this subsection, we illustrate how to use the proposed framework to
compare various existing user association and resource allocation
strategies in terms of network power consumption.

The first strategy is the proposed algorithm in this paper, where all
$2^{15}$ patterns are considered in the candidate pattern set.

The second is the \emph{Reuse-1} scheme \cite{Pollakis2012,
  Cavalcante2014}, which can be cast into the proposed framework by
restricting the candidate pattern to a single reuse-1 pattern (see
Section \ref{sec:noICIC}).

The third one is the \emph{Pre-selected feature patterns} scheme. The
idea is to group pico BSs within one macro cell into one cluster when
formulating interference patterns (see discussions in Section
\ref{sec:cell-clust-activ}). In such way, the number of possible
patterns has been reduced from $2^{15}$ to $2^{6}$. We further
restrict the candidate patterns to the following \emph{four} by
switching on: \{p1,p2,p3\}, \{1,p2,p3\}, \{2,p1,p3\}, and \{3,p1,p2\},
where p1, p2, p3 denote the pico clusters within cells 1, 2, and 3,
respectively. This was a suggested feature pattern set in
\cite{Kuang2014a} achieving nearly optimal rate utility
maximization. In addition to these four patterns, we add 15 patterns,
each one activating one single cell, in order to increase the
granularity of interference characterization.

The fourth strategy is to separate the user association from the joint
optimization and apply the simple range expansion rule.
In the evaluation, we set the macro bias to zero, and the same bias for all the pico BSs,
choosing from 0, 10, 20, 30 and 40 dB.\footnote{In the current LTE
  networks, the maximum bias for pico range expansion is typically
  restricted to 15 dB. Too aggressive bias will potentially cause the
  control channel failure. In this paper, we do not consider this
  restriction when investigating the full potential of the range
  expansion scheme. Our consideration can be justified by assuming a
  split between U-plane and C-plane, and there exists some control
  channel protection mechanism, e.g. control channels of pico and
  macro are deployed on the orthogonal resources. } All possible
patterns are included in the resource allocation.

\begin{figure}%
\centering
\includegraphics[width=3.2In]{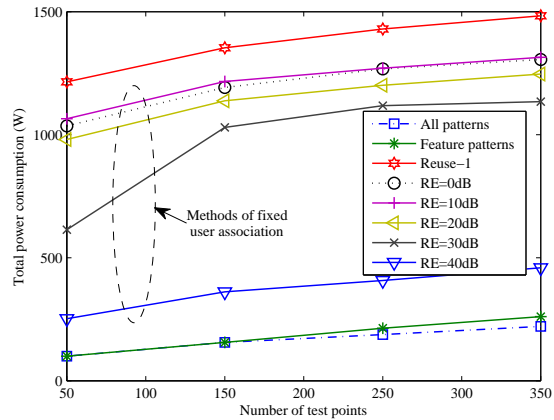}
\caption{Network power consumption of different schemes. All test
  points have the same rate requirement $d_k = 200$kbit/s, and uniformly distributed in the network. The results are
  averaged over ten realizations for each given number of test points.
}\label{fig:compare}
\end{figure}

The results are given in Fig.~\ref{fig:compare}. As shown, the proposed algorithm considering all patterns achieves the
minimum power consumption for all the given test-point-cases. We can
use it as a benchmark for quantifying other strategies where the
patterns are somehow restricted or resource optimization is decoupled
from the user association (by RE rules).

The feature-pattern scheme achieves almost the same power saving
performance as using all patterns. This is because there is little
loss in the user achievable rates by characterizing interference using
this feature pattern set, as shown in \cite{Kuang2014a}. When it is
used for activating cells to satisfy the user rate demand, it is not
surprising to achieve close-to-benchmark performance.


The existing strategy based on reuse-1, on the other hand, achieves
the worst power saving performance among all the methods in
comparison. This is because the interference coupling has not be taken
into account when (de)activating cells.
It completely neglects the fact that muting some BSs can reduce the
interference and hence increase the user rate. Therefore, the
resulting BS activation and user association decisions are highly
sub-optimal. For example, as shown in Fig.~\ref{fig:compare}, the
proposed algorithms (all patterns and feature patterns) only need
$12\%$ of the power consumption of the reuse-1 scheme to support $150$
test points with $200$kbit/s.

We can also observe from Fig.~\ref{fig:compare} that the methods with
the fixed RE association achieve the performance between the benchmark
and the worst reuse-1 method. Moreover, by increasing the pico biases
from 0 dB to 40 dB, more and more test points are offloaded onto the
pico cells, increasing the opportunity of deactivating macro
cells. Hence, it reduces the power consumption. However, we still
observe a considerable performance loss even a large bias of 40 dB is
used for pico cells, compared to the benchmark. This is because all
pico BSs have to use the same bias value. We will make this point
clearer by computing cell-specific biases in the next subsection.

\begin{table*}[t]
\renewcommand{\arraystretch}{1.0}
\caption{Mapping jointly optimized solution to cell-specific biases (in dB). Case I: 50 test points, $d_k=0.2$Mbit/s; Case II: 150 test points, $d_k=0.2$Mbit/s; Case III: 50 test points, $d_k=1.0$Mbit/s. Error is defined as the ratio of number of wrong association to the number of test points. }
\label{tab:optimalBias}
\centering
\begin{tabular}{|c | c | c | c | c|c|c | c | c | c | c|c|c|c|c|}
\hline
Cell index & 1,2,3 & 4 & 5 & 6 & 7 & 8 & 9 & 10 & 11 & 12 & 13 & 14 & 15 &  \\
\hline
\hline
Case I  & 0 & 0 & 40.7 & 0 & 25.2 & 4.8 & 40.6 & 42.6 & 0 & 32.1 & 0 & 0 & 38.3 & 0\% error \\
\hline
Case II & 0 & 0 & 41.0& 36.8& 25 & 35 & 40.7 & 40.6 & 0 & 34.8 & 33.1 & 33.4 & 35.9 & 1.33\% error\\
\hline
Case III & 0 & 28.1 & 40.9 & 36.4 & 31.1 & 33.1 & 40.6 & 42.6 & 0 & 32.1 & 35.7 & 5.6 & 38.3 & 0\% error\\
\hline
\end{tabular}
\end{table*}

%

\subsection{Cell-specific bias}
\label{sec:simuBias}

In Table~\ref{tab:optimalBias}, we show the cell-specific bias
obtained from the jointly optimized solution for a typical network
realization using the algorithm described in Section
\ref{sec:mapp-optim-assoc}. As can be seen, the minimization of
network power consumption requires different bias for different
cells. This is in sharp contrast to the conclusions derived in
\cite{Ye2013} from the perspective of load balancing, where the same
bias per-tier resulted in almost no performance loss in the network
rate utility. This is because biasing for energy saving is targeted at
a different goal from load balancing. In order to deactivate some BSs,
biasing here is used as a mechanism to concentrate users to a small
set of cells.

\begin{figure}
\centering
\includegraphics[width=3.2In]{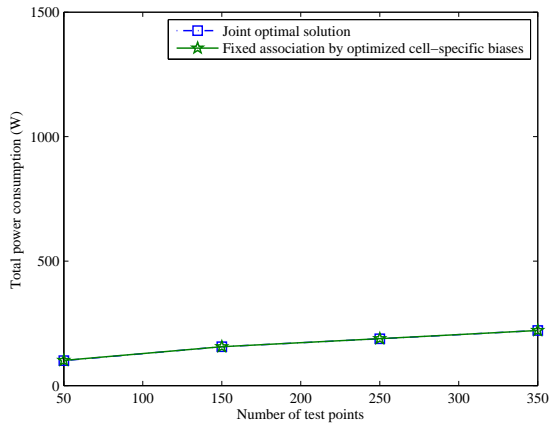}
\caption{Comparison of the jointly optimized solution with the fixed
  association using optimized cell-specific bias. All test points have
  the same rate requirement $d_k = 200$kbit/s, and
  uniformly distributed in the network. The results are averaged over
  ten realizations for each given number of test points.}
\label{fig:compare2}
\end{figure}

By using the derived cell-specific bias from the jointly optimized
solution, we again solve the network power minimization problem of
\eqref{eq:5} with fixed user association, and report the results in
Fig.~\ref{fig:compare2}. As shown, the previous gap in
Fig.~\ref{fig:compare} between the benchmark and the scheme with fixed
RE association is now closed.

\subsection{Modeling aspects and practical implementation of the solution}
\label{sec:modelingIssue}

During the relatively long decision period considered in this paper,
users could join the network and then leave after being served. Hence,
we adopt test points as an abstract concept to represent demands of
users in a given region, as in \cite{Cavalcante2014}. The test points
can be chosen from typical user locations, or we can simply partition
the geographic region into pixels, within each pixel radio propagation
being considered uniform and then each pixel becomes one test point.

In the proposed model, the demand is represented by a minimum required average
rate $d_k$, similar to that in \cite{Pollakis2012,
  Cavalcante2014}. Generally speaking, the demand can be calculated
from the QoS requirement of users at the test point. For example, at
point $k$, we can assume that file transfer requests arrive following
a Poisson point process with the mean arrival rate $\lambda_k$ in
$\textrm{s}^{-1}$ and the exponentially distributed file size with
mean $L_k$ in bits, resulting in an average traffic load $\Omega_k =
\lambda_k L_k$ in bit/s \cite{Son2011a}. The file transfer requests at
the same test point are served according to a first-come-first-served
policy. Hence, each test point has effectively an M/M/1 queue. Suppose
the QoS of users at test point $k$ requires the average file sojourn
time (or response time) is not greater than a given value $\tau_k$ (in
second), which can be expressed as $\frac{1}{R_k/L_k - \lambda_k} \leq
\tau_k$ (see \cite{Nelson, ZhuangEnergy}), where $R_k$ is the average deliverable
data rate of point $k$ as given in \eqref{RateCombined}, and $R_k/L_k$
is the mean service rate in $\textrm{s}^{-1}$. This translates into
the average data rate constraint $R_k \geq d_k$, as expressed in
\eqref{conQoS}, with $d_k = L_k/\tau_k + \Omega_k$.

The proposed algorithms need the knowledge of user demand $\{d_k,
\forall k\}$ and deliverable rate $\{r_{kbi}, \forall k,b,i\}$ at a
central controller where resource management is executed. Location
information is also required to identify which test point a user
belongs to, which can be obtained by standard positioning methods. The
serving BSs estimate/predict traffic pattern based on the traffic
aggregation from all users within each test point
\cite{Cavalcante2014}. Based on the estimated traffic information and
QoS, the serving BS calculates the user demand of the associated test
points and forwards this information to the central controller.

The deliverable rate $r_{kbi}$ is calculated at the central
controller. To facilitate this calculation, each BS should forward the
channel gains between itself and all test points to the central
controller. Note that channel information is routinely measured by the
BSs in the current mobile network standards, either relying on
uplink-downlink reciprocity or feedback from users. Since the resource
management is adapted at a slow timescale, the BSs can only forward
wideband channel coefficients $G_{bk}$ in (2) and ignore the
frequency-selective channel coefficients to reduce the overhead.

Once the central controller calculates the pattern resource parameter
$\boldsymbol{\pi}$ and user resource parameter $\boldsymbol{\alpha}$
by proposed algorithms, it informs all BSs of these decision
variables.  Each BS is only allowed to have access to certain fraction
of system bandwidth as specified by $\pi_i$ (see
Fig.\ref{fig1_illustration}). Regarding the user association, the
central controller decides which test point a user should belong to
based on the location information.
File transmission requests within a given test point $k$ are routed to
BS $b$ by the central controller according to the association
decision. Then BS $b$ allocates certain fraction of bandwidth under
pattern $i$ to serve the transmission, as specified by
$\alpha_{kbi}$. On top of this adaptation, each BS can perform
individual channel-aware scheduling for its associated users among the
agreed spectrum in a more frequent manner to respond to fast fading
channel fluctuations.

\section{Conclusion}
\label{sec:conclusion}

Interference coupling in heterogeneous networks introduces the
inherent non-convexity to the multi-cell resource optimization
problem, hindering the development of effective solutions. A new
framework based on multi-pattern formulation has been proposed in this
paper to study the energy efficient strategy for joint cell
activation, user association and channel allocation. One key feature
of this interference pattern formulation is that the patterns remain
fixed and independent of the optimization process. This creates a
favorable opportunity for convex formulation while still taking
interference coupling into account. By grouping weakly
mutual-interfering cells when formulating possible interference
patterns in the network, and then allocating resources among these
patterns, we arrive at an optimization problem with controllable
complexity.
A tailored algorithm has been proposed based on the reweighted
$\ell_1$ minimization and the cutting plane method in the dual domain by
exploiting the problem structure, resulting in significant complexity
saving. Relying on this algorithm, a benchmark involving all $2^B$
possible patterns in the optimization has been derived to quantify the
existing solutions with restricted patterns. Numerical results have
demonstrated a high power saving by the proposed strategy. In contrast
to previous studies on load balancing, per-tier biasing rule is not
optimal for energy saving investigated in this paper.

\appendix[Proof of Proposition 1]

By letting $\alpha_{kbi} = \pi_i\theta_{kbi}$, the original problem can be equivalently rewritten as
\begin{IEEEeqnarray}{rCl}\label{P1_rateConstrained_changeVairable}
\IEEEyesnumber\IEEEyessubnumber*
  \displaystyle\mathop{\minimize}_{\boldsymbol{\theta},\boldsymbol{\pi}}
  \quad
  &&    P^{\text{tot}}= \sum_{b\in \mathcal{B}} \left[ (1-q_b)\rho_b P_b^\textrm{OP} +
    q_b |\rho_b|_0 P_b^\textrm{OP} \right]  \label{prop1_obj} \IEEEeqnarraynumspace\\
  \text{subject to} \quad && \rho_b =  \sum_{i\in\mathcal{I}} \pi_i
  \sum_{k\in\mathcal{K}} \theta_{kbi}, \  \forall b \label{prop1_cons_rho} \\
  && \sum_{i \in \mathcal{I}} \pi_i \sum_{b \in \mathcal{B}}  \theta_{kbi} r_{kbi} \geq d_k, \ \forall k  \label{prop1_conQoS} \\
  && \sum_{k \in \mathcal{K}} \theta_{kbi} \leq 1, \forall b,\ \forall i    \label{prop1_const_BS allo}\\
  && \sum_{i \in \mathcal{I}} \pi_i = 1  \label{prop1_cons_pi} \\
  && \pi_i \geq 0, \ \forall i, \quad \theta_{kbi} \geq 0, \ \forall
  k,b,i  \label{prop1_con_nonnegative}
\end{IEEEeqnarray}
In the following, we show that if an optimal solution
$(\boldsymbol{\theta}^\star, \boldsymbol{\pi}^\star)$ exists we can
then obtain the same optimal objective with
$(\boldsymbol{\theta}^\star, \boldsymbol{\pi'})$ where
$\boldsymbol{\pi'}$ only has $K+B+1$ nonzero entries out of $
|\mathcal{I}|$ entries.

We first define $\mathbf{t}_i=(t_{1i},\cdots,t_{bi},\cdots, t_{Bi})^T$
with $t_{bi} = \sum_{k\in\mathcal{K}} \theta_{kbi}^\star$, and
$\mathbf{R}_i = (R_{1i},\cdots, R_{ki}, \cdots$, $R_{Ki})^T$ with
$R_{ki} = \sum_{b \in \mathcal{B}} \theta_{kbi}^\star r_{kbi}$. Then
define $\boldsymbol{\rho} = (\rho_1,\cdots, \rho_B)^T$ and $\mathbf{d}
= (d_1, \cdots, d_K)^T$. According to \eqref{prop1_cons_rho} and
\eqref{prop1_conQoS} (note that \eqref{prop1_conQoS} must achieve
equality at the optimum, otherwise the objective in \eqref{prop1_obj}
can be further reduced), the vector $(\boldsymbol{\rho}^T,
\mathbf{d}^T)^T = \sum_i \pi_i (\mathbf{t}_i^T, \mathbf{R}_i^T)^T$,
i.e., a convex combination of vectors $(\mathbf{t}_i^T,
\mathbf{R}_i^T)^T, \forall i\in \mathcal{I}$, with $\pi_i$ as
coefficients. By Caratheodory's Theorem, $(\boldsymbol{\rho}^T,
\mathbf{d}^T)^T$ can be represented by at most $K+B+1$ of those
vectors. Denoting the resulting coefficients by
$\boldsymbol{\pi}^\prime$, we prove the proposition.


\begin{thebibliography}{10}
\providecommand{\url}[1]{#1}
\csname url@samestyle\endcsname
\providecommand{\newblock}{\relax}
\providecommand{\bibinfo}[2]{#2}
\providecommand{\BIBentrySTDinterwordspacing}{\spaceskip=0pt\relax}
\providecommand{\BIBentryALTinterwordstretchfactor}{4}
\providecommand{\BIBentryALTinterwordspacing}{\spaceskip=\fontdimen2\font plus
\BIBentryALTinterwordstretchfactor\fontdimen3\font minus
  \fontdimen4\font\relax}
\providecommand{\BIBforeignlanguage}[2]{{%
\expandafter\ifx\csname l@#1\endcsname\relax
\typeout{** WARNING: IEEEtran.bst: No hyphenation pattern has been}%
\typeout{** loaded for the language `#1'. Using the pattern for}%
\typeout{** the default language instead.}%
\else
\language=\csname l@#1\endcsname
\fi
#2}}
\providecommand{\BIBdecl}{\relax}
\BIBdecl

\bibitem{Bhushan2014}
N.~Bhushan, J.~Li, D.~Malladi, R.~Gilmore, D.~Brenner, A.~Damnjanovic,
  R.~Sukhavasi, C.~Patel, and S.~Geirhofer, ``Network densification: the
  dominant theme for wireless evolution into 5G,'' \emph{{IEEE} Commun. Mag.},
  vol.~52, no.~2, pp. 82--89, 2014.

\bibitem{Son2011a}
K.~Son, H.~Kim, Y.~Yi, and B.~Krishnamachari, ``Base station operation and user
  association mechanisms for energy-delay tradeoffs in green cellular
  networks,'' \emph{{IEEE} J. Sel. Areas Commun.}, vol.~29, no.~8, pp.
  1525--1536, 2011.

\bibitem{Pollakis2012}
E.~Pollakis, R.~Cavalcante, and S.~Stanczak, ``Base station selection for
  energy efficient network operation with the majorization-minimization
  algorithm,'' in \emph{Signal Processing Advances in Wireless Communications
  (SPAWC), 2012 IEEE 13th International Workshop on}, 2012, pp. 219--223.

\bibitem{Cavalcante2014} R.~L.~G. {Cavalcante}, S.~{Sta{\'n}czak},
  M.~{Schubert}, A.~{Eisenbl{\"a}tter}, and U.~{T{\"u}rke}, ``Toward
  Energy-Efficient 5G Wireless Communications Technologies: Tools for
  decoupling the scaling of networks from the growth of operating
  power,'' \emph{IEEE Signal Processing Magazine}, vol.~31, no.~6,
  pp.~24--34, 2014.

\bibitem{Kim2013a}
S.~Kim, S.~Choi, and B.~G. Lee, ``A joint algorithm for base station operation
  and user association in heterogeneous networks,'' \emph{{IEEE} Commun.
  Lett.}, vol.~17, no.~8, pp. 1552--1555, 2013.

\bibitem{Su2013}
L.~Su, C.~Yang, Z.~Xu, and A.~Molisch, ``Energy-efficient downlink transmission
  with base station closing in small cell networks,'' in \emph{Acoustics,
  Speech and Signal Processing (ICASSP), 2013 IEEE International Conference
  on}, 2013, pp. 4784--4788.

\bibitem{YuanMingTWC}
Y.~Shi, J.~Zhang, and K.~Letaief, "Group sparse beamforming for green cloud-RAN", \emph{IEEE Trans. Wireless Commun.},vol.~13, no.~5, pp.~2809--2823, 2014.

\bibitem{Liao2013}
W.-C. Liao, M.~Hong, Y.-F. Liu, and Z.-Q. Luo, ``Base station activation and
  linear transceiver design for optimal resource management in heterogeneous
  networks,'' \emph{IEEE Trans. Signal Process.}, vol.~62, no.~15,
  pp.~3939--3952, 2014.

\bibitem{Kuang2014a}
Q.~Kuang, W.~Utschick, and A.~Dotzler, ``Optimal joint user association and
  resource allocation in heterogeneous networks via sparsity pursuit,''
  \emph{arXiv:1408.5091}, Aug. 2014.

\bibitem{Zhuangtobepublished}
\BIBentryALTinterwordspacing
B.~Zhuang, D.~Guo, and M.~Honig, ``Traffic-driven spectrum allocation in
  heterogeneous networks,'' \emph{{IEEE} J. Sel. Areas Commun.}, vol.~33, no.~10, pp. 2027--2038, May 2015.
\BIBentrySTDinterwordspacing

\bibitem{Candes2008}
E.~J. Candes, M.~B. Wakin, and S.~P. Boyd, ``Enhancing sparsity by reweighted 1
  minimization,'' \emph{Journal of Fourier analysis and applications}, vol.~14,
  no. 5-6, pp. 877--905, 2008.

\bibitem{Damnjanovic2011}
A.~Damnjanovic, J.~Montojo, Y.~Wei, T.~Ji, T.~Luo, M.~Vajapeyam, T.~Yoo,
  O.~Song, and D.~Malladi, ``A survey on 3GPP heterogeneous networks,''
  \emph{IEEE Wireless Communications}, vol.~18, no.~3, pp. 10--21, 2011.

\bibitem{Ye2013}
\BIBentryALTinterwordspacing
Q.~Ye, B.~Rong, Y.~Chen, M.~Al-Shalash, C.~Caramanis, and J.~Andrews, ``User
  association for load balancing in heterogeneous cellular networks,''
  \emph{{IEEE} Trans. Wireless Commun.}, vol.~12, no.~6, pp. 2706--2716, 2013.


\bibitem{Niu2010}
Z.~Niu, Y.~Wu, J.~Gong, and Z.~Yang, ``Cell zooming for cost-efficient green
  cellular networks,'' \emph{{IEEE} Commun. Mag.}, vol.~48, no.~11, pp. 74--79,
  2010.

\bibitem{Nesterov1994}
Y.~Nesterov, A.~Nemirovskii, and Y.~Ye, \emph{Interior-point polynomial
  algorithms in convex programming}.\hskip 1em plus 0.5em minus 0.4em\relax
  SIAM, 1994, vol.~13.

\bibitem{Boyd2004}
S.~Boyd and L.~Vandenberghe, \emph{Convex Optimization}.\hskip 1em plus 0.5em
  minus 0.4em\relax Cambridge University Press, New York, USA, 2004.

\bibitem{Bazaraa2013}
M.~S. Bazaraa, H.~D. Sherali, and C.~M. Shetty, \emph{Nonlinear programming:
  theory and algorithms}, 3rd~ed.\hskip 1em plus 0.5em minus 0.4em\relax New
  York: Wiley-Interscience, 2006.

\bibitem{Kuang2012}
Q.~Kuang, J.~Speidel, and H.~Droste, ``Joint base-station association, channel
  assignment, beamforming and power control in heterogeneous networks,'' in
  \emph{Vehicular Technology Conference (VTC Spring), 2012 IEEE 75th}, 2012,
  pp. 1--5.

\bibitem{Lipp2014}
T.~Lipp and S.~Boyd, ``Variations and extensions of the convex-concave
  procedure,'' 2014.

\bibitem{3GPP214}
3GPP, ``Evolved universal terrestrial radio access (e-utra); physical layer;
  measurements (ts 36.214),'' April 2011.

\bibitem{Madan2010}
R.~Madan, J.~Borran, A.~Sampath, N.~Bhushan, A.~Khandekar, and T.~Ji, ``Cell
  association and interference coordination in heterogeneous LTE-A cellular
  networks,'' \emph{IEEE Journal on Selected Areas in Communications}, vol.~28,
  no.~9, pp. 1479--1489, 2010.

\bibitem{3GPP2010}
3GPP, ``Further advancements for e-utra physical layer aspects (tr 36.814),''
  vol. v9.0.0, 2010.

\bibitem{Fehske2009}
\BIBentryALTinterwordspacing
A.~Fehske, F.~Richter, and G.~Fettweis, ``Energy efficiency improvements
  through micro sites in cellular mobile radio networks,'' in \emph{GLOBECOM
  Workshops, 2009 IEEE}, 2009, pp. 1--5.


\bibitem{kuangISWCS}
Q.~Kuang, ``Joint user association and reuse pattern selection in heterogeneous
  networks,'' in \emph{Proc. IEEE Int. Symp. on Wireless
  Communication Systems (ISWCS)}, Spain, August 2014.

\bibitem{Nelson}
  R.Nelson, \emph{Probability, stochastic process, and queueing theory: the mathematics of computer performance modeling}, Springer, 1995.

\bibitem{ZhuangEnergy}
B.~Zhuang, D.~Guo, and M.~Honig, ``Energy-efficient cell activation, user association, and spectrum allocation in heterogeneous networks,'' \emph{{IEEE} J. Sel. Areas Commun.}, to appear, available: \emph{arXiv:1509.04805}, Sep. 2015.

\bibitem{ShixinTWC2015}
S.~Luo, R.~Zhang, and T.~J.~Lim, "Downlink and uplink energy minimization through user association and beamforming in C-RAN." \emph{IEEE Trans. Wireless Commun.}, vol.~14, no.~1, pp.~494--508, 2015.

\bibitem{kuangICASSP16}
Q.~Kuang, X.~Yu, and W.~Utschick, "Network topology adaptation and interference coordination for energy saving in heterogeneous networks", accepted at \emph{ICASSP 2016}, available : \emph{arXiv:1511.06888}.

\bibitem{Lanckriet2009}
G.~R.~Lanckriet, and B.~K.~Sriperumbudur, "On the convergence of the concave-convex procedure." \emph{Advances in neural information processing systems}, pp. 1759 -- 1767. 2009.

\bibitem{gurobi}
Gurobi Optimization, Inc., ``Gurobi optimizer reference manual,'' http://www.gurobi.com, 2015.

\end{thebibliography}
\end{document}